\documentclass[prl,twocolumn,preprintnumbers,amsmath,amssymb,floatfix,longbibliography,superscriptaddress]{revtex4-2}
\usepackage{graphicx}
\usepackage{epstopdf}
\usepackage{psfrag}
\usepackage{hyperref}
\usepackage{multirow}
\usepackage[sort&compress]{natbib}
\usepackage{amssymb}
\usepackage{amsmath}
\usepackage{amsthm}
\usepackage{bbold}
\usepackage{bbm}
\usepackage{enumerate}
\usepackage{textcomp}
\usepackage{verbatim}
\usepackage{soul}
\usepackage{ulem}
\usepackage[caption=false]{subfig}
\usepackage{orcidlink}

\newcommand{\vphi}{\varphi}

\newcommand{\vare}{\varepsilon}

\newcommand{\rmi}{{\rm i}}

\newcommand{\beq}{\begin{equation}\begin{aligned}}
\newcommand{\eeq}{\end{aligned}\end{equation}}

\usepackage{xcolor}

\definecolor{AC}{rgb}{1, 0.2, 0.7}
\definecolor{SD}{rgb}{1,0,1}

\graphicspath{{./}{./figs/}}

\begin{document}

\hypersetup{pdftitle={title}}
\title{Simulating Holographic Conformal Field Theories on Hyperbolic Lattices}

\author{Santanu Dey\,\orcidlink{0000-0003-2125-6163}}
\email{santanu@ualberta.ca}
\affiliation{Department of Physics, University of Alberta, Edmonton, Alberta T6G 2E1, Canada}
\affiliation{Theoretical Physics Institute, University of Alberta, Edmonton, Alberta T6G 2E1, Canada}

\author{Anffany Chen\,\orcidlink{0000-0002-0926-5801}}
\email{anffany@ualberta.ca}
\affiliation{Department of Physics, University of Alberta, Edmonton, Alberta T6G 2E1, Canada}
\affiliation{Theoretical Physics Institute, University of Alberta, Edmonton, Alberta T6G 2E1, Canada}

 \author{Pablo Basteiro}
 \affiliation{Institute for Theoretical Physics and Astrophysics, Julius Maximilians University W\"urzburg, Am Hubland, 97074 Würzburg, Germany}
 \affiliation{W\"urzburg-Dresden Excellence Cluster ct.qmat, Julius Maximilians University W\"urzburg, Am Hubland, 97074 Würzburg, Germany}

\author{Alexander Fritzsche}
\affiliation{Institute for Theoretical Physics and Astrophysics, Julius Maximilians University W\"urzburg, Am Hubland, 97074 Würzburg, Germany}
 \affiliation{W\"urzburg-Dresden Excellence Cluster ct.qmat, Julius Maximilians University W\"urzburg, Am Hubland, 97074 Würzburg, Germany}
 \affiliation{Institut f\"ur Physik, Universit\"at Rostock, 18059 Rostock, Germany}

\author{Martin Greiter\,\orcidlink{0000-0003-2008-4013}}
\affiliation{Institute for Theoretical Physics and Astrophysics, Julius Maximilians University W\"urzburg, Am Hubland, 97074 Würzburg, Germany}
 \affiliation{W\"urzburg-Dresden Excellence Cluster ct.qmat, Julius Maximilians University W\"urzburg, Am Hubland, 97074 Würzburg, Germany}

\author{Matthias Kaminski\,\orcidlink{0000-0002-0261-6031}}
 \affiliation{Department of Physics and Astronomy, University of Alabama, Tuscaloosa, AL
35487, USA}

\author{Patrick M. Lenggenhager\,\orcidlink{0000-0001-6746-1387}}
 \affiliation{W\"urzburg-Dresden Excellence Cluster ct.qmat, Julius Maximilians University W\"urzburg, Am Hubland, 97074 Würzburg, Germany}
\affiliation{Department of Physics, University of Zurich, 8057 Zurich, Switzerland}
\affiliation{Condensed Matter Theory Group, Paul Scherrer Institute, 5232 Villigen, Switzerland}
\affiliation{Institute for Theoretical Physics, ETH Zurich, 8093 Zurich, Switzerland}
\affiliation{Max Planck Institute for the Physics of Complex Systems, Nöthnitzer Str. 38, 01187 Dresden, Germany}

\author{Ren\'e Meyer\,\orcidlink{0000-0002-6607-8199}}
\affiliation{Institute for Theoretical Physics and Astrophysics, Julius Maximilians University W\"urzburg, Am Hubland, 97074 Würzburg, Germany}
 \affiliation{W\"urzburg-Dresden Excellence Cluster ct.qmat, Julius Maximilians University W\"urzburg, Am Hubland, 97074 Würzburg, Germany}

 \author{Riccardo Sorbello}
\affiliation{Institute for Theoretical Physics and Astrophysics, Julius Maximilians University W\"urzburg, Am Hubland, 97074 Würzburg, Germany}
 \affiliation{W\"urzburg-Dresden Excellence Cluster ct.qmat, Julius Maximilians University W\"urzburg, Am Hubland, 97074 Würzburg, Germany}

 \author{Alexander Stegmaier\,\orcidlink{0000-0002-8864-5182}}
\affiliation{Institute for Theoretical Physics and Astrophysics, Julius Maximilians University W\"urzburg, Am Hubland, 97074 Würzburg, Germany}
 \affiliation{W\"urzburg-Dresden Excellence Cluster ct.qmat, Julius Maximilians University W\"urzburg, Am Hubland, 97074 Würzburg, Germany}

\author{Ronny Thomale\,\orcidlink{0000-0002-3979-8836}}
\affiliation{Institute for Theoretical Physics and Astrophysics, Julius Maximilians University W\"urzburg, Am Hubland, 97074 Würzburg, Germany}
 \affiliation{W\"urzburg-Dresden Excellence Cluster ct.qmat, Julius Maximilians University W\"urzburg, Am Hubland, 97074 Würzburg, Germany}
  
\author{Johanna Erdmenger  \orcidlink{0000-0003-4776-4326}}
\affiliation{Institute for Theoretical Physics and Astrophysics, Julius Maximilians University W\"urzburg, Am Hubland, 97074 Würzburg, Germany}
 \affiliation{W\"urzburg-Dresden Excellence Cluster ct.qmat, Julius Maximilians University W\"urzburg, Am Hubland, 97074 Würzburg, Germany}

\author{Igor Boettcher\,\orcidlink{0000-0002-1634-4022}}
 \affiliation{Department of Physics, University of Alberta, Edmonton, Alberta T6G 2E1, Canada}
 \affiliation{Theoretical Physics Institute, University of Alberta, Edmonton, Alberta T6G 2E1, Canada}

\begin{abstract}
We demonstrate how table-top settings combining hyperbolic lattices with nonlinear dynamics universally encode aspects of the bulk-boundary-correspondence between gravity in anti-de-Sitter (AdS) space and conformal field theory (CFT).
Our concrete and broadly applicable holographic toy model simulates gravitational self-interactions in the bulk and features an emergent CFT with nontrivial correlations on the boundary.
We measure the CFT data contained in the two- and three-point functions and clarify how a thermal CFT is simulated through an effective black hole geometry. As a concrete example, we propose and simulate an experimentally feasible protocol to measure the holographic CFT using electrical circuits.
\end{abstract}

\maketitle

The holographic principle as realized by the AdS/CFT correspondence postulates a deep connection between two of the most intriguing, yet unfathomable phenomena in modern physics \cite{maldacena1999large,witten1998,gubser98,ammon15}: scale-invariance close to second-order phase transitions, described by conformal field theories (CFTs) \cite{
bigyellow97}, and quantum gravity in curved spacetimes. The conjecture states that gravity in a negatively curved space is dual to a CFT on the boundary of that space. Furthermore, the presence of a black hole in the bulk results in a thermal CFT, with the temperature given by the Hawking temperature. While numerous calculations indicate the validity of the correspondence, an experimental verification is complicated  in particular by the need to observe or simulate graviton-graviton interactions, since  only these give rise to nonvanishing three- or four-point boundary correlation functions, and nontrivial CFT data.

The goal of this work is to construct a concrete holographic toy model for the AdS/CFT correspondence that can be realized in the laboratory. We show that, with the right four ingredients, a large class of realistic and experimentally feasible low-energy models features holo\-graphy through conformal boundary correlations. As a concrete example, we demonstrate how electrical circuits with nonlinear circuit elements can achieve this milestone, but our universal model is applicable to other experimental platforms and constitutes a theoretically intriguing model in itself. With the setup, we are able to emulate three-point functions, characterize the CFT data, and clarify how thermal effects are incorporated.

The four ingredients of our holographic toy model are: (i) a lattice realization of AdS space through hyperbolic lattices \cite{kollar2019hyperbolic,kollar2019line,PhysRevX.10.011009,Yu2020,PhysRevD.102.034511,PhysRevA.102.032208,Jahn2020,brower2021lattice,Zhu:2021,maciejko2020hyperbolic,Bienias2022,Maciejko2022,Boettcher2022,Stegmaier2021,zhang2022observation,Liu2022,PhysRevD.105.114503,Lenggenhager2021,cheng2022band,PhysRevE.106.034114,PhysRevB.106.155146,10.21468/SciPostPhys.13.5.103,Urwyler2022,chen2023hyperbolic,zhang2023hyperbolic,gluscevich2023dynamic,PhysRevB.107.125302,PhysRevLett.130.091604,PhysRevB.107.165145,chen2023adscft,PhysRevB.107.184201,gluscevich2023magnetic,petermann2023eigenmodes,PhysRevB.108.035154,Chen2023Symmetry,curtis2023absence,shankar2023hyperbolic,schrauth2023hypertiling,chen2023anderson,PhysRevLett.131.226401,li2023anderson,10.21468/SciPostPhys.15.5.218,PhysRevB.109.L041109,tummuru2023hyperbolic}, now routinely implemented in circuit quantum electrodynamics and coplanar waveguides \cite{kollar2019hyperbolic,fleury2024}, topoelectrical circuits \cite{Lenggenhager2021,zhang2022observation,chen2023hyperbolic,zhang2023hyperbolic}, topological photonics \cite{huang2024hyperbolic}, and mechanical elastic lattices \cite{patino2024hyperbolic}; (ii) nonlinear dynamical equations to emulate gravitational self-interactions, realized by local qubit-photon interactions \cite{houck2012chip,PhysRevX.9.011021,Bienias2022}, nonlinear or active circuit elements \cite{PhysRevLett.114.173902,PhysRevX.5.021031,lee2018topolectrical,kotwal2021active,PhysRevResearch.5.L012041}, nonlinear optics \cite{RevModPhys.91.015006,PhysRevA.108.040101}, or spring hardening; (iii) effective black hole geometries to emulate temperature by using type-II hyperbolic lattices \cite{banados92,carlip95,Keski1999,chen2023adscft}; (iv) a theoretical framework and experimental protocol to compute and measure boundary correlation functions. We demonstrate that with (i)-(iv) even classical platforms are holographic, corresponding to the limit where weakly-coupled semiclassical bulk gravity is dual to a strongly coupled CFT, as reviewed in \cite{ammon15}.

\begin{figure*}[t!]
\includegraphics[width=\linewidth]{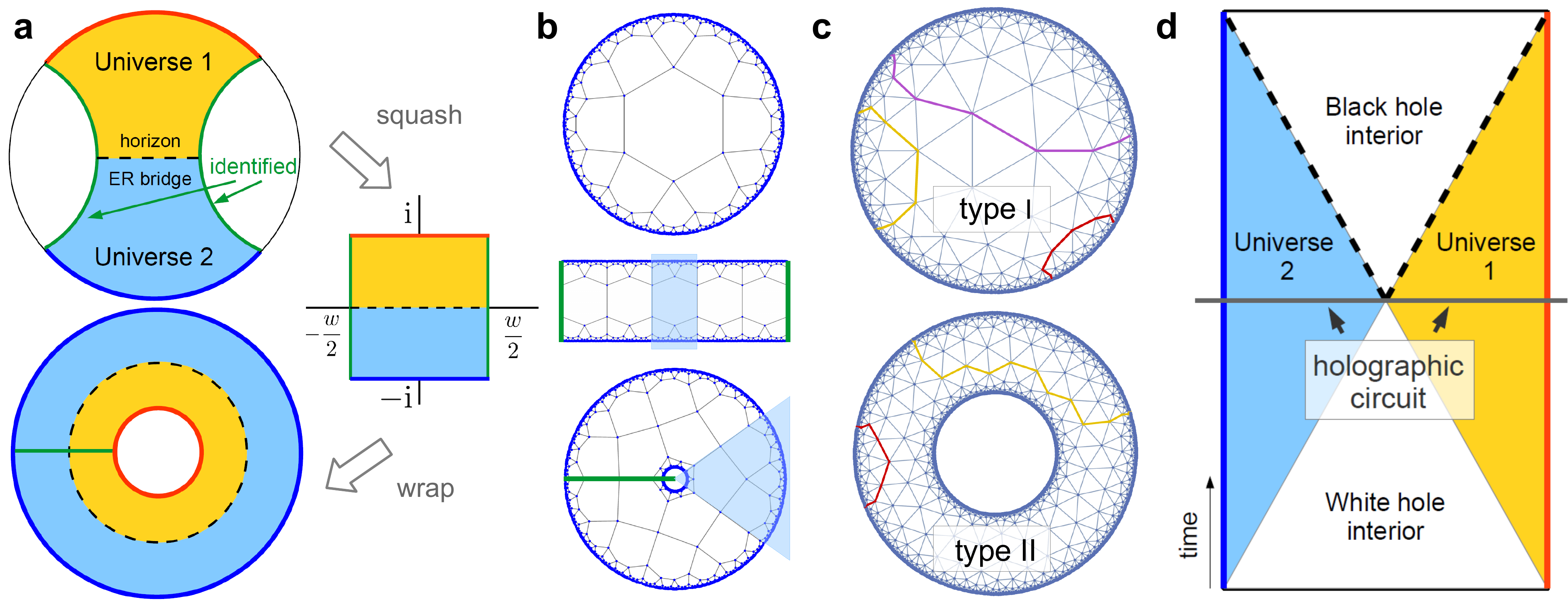}
\caption{\textbf{Hyperbolic flakes and black hole geometry.} \textbf{(a)} Starting from a Poincar\'{e} disk (type-I geometry, coordinates $z=re^{\rmi \theta}$), we create a black hole by identifying points on two geodesics (shown in green). Equivalently, we squash the disk to form a strip of width $w$ and then wrap it into a ring (type-II geometry, coordinates $\hat{z}=\hat{r}e^{\rmi \theta}$). This identification creates two separate Universes, 1 and 2, each with a holographic boundary, connected through an Einstein--Rosen (ER) bridge (dashed line) \cite{MaldacenaSusskind}. This line corresponds to the bifurcation point of an eternal black hole horizon. \textbf{(b)} Applying the squash-and-wrap procedure to a hyperbolic $\{p,q\}$-flake creates a type-II flake, which emulates a black hole with Hawking temperature $T=w/8\pi\ell$. Herein, $w=kP$, with $k$ an integer representing the number of repeated cells in the strip ($k=5$ in the plot), and $P=P(p,q)$ a constant. \textbf{(c)} Type-I and -II hyperbolic flakes of a $\{3,7\}$ tessellation with some graph-shortest-distance paths connecting boundary sites. These paths are pivotal for the boundary-boundary correlations. \textbf{(d)} Schematic Penrose diagram with space and time as x- and y-axis, respectively. The hyperbolic lattice is located on a slice of constant time (gray line).}
\label{FigBTZ}
\end{figure*}

Our work considerably extends previous works on holographic aspects of hyperbolic lattices. Power-law scaling of two-point boundary correlators has been demonstrated at zero temperature in the important works \cite{brower2021lattice,PhysRevD.102.034511,chen2023adscft}, together with the massless contact-limit of the four-point function \cite{brower2021lattice}, the introduction of black holes through type-II lattices \cite{chen2023adscft}, and the emergence of the Ryu--Takayanagi formula \cite{chen2023adscft}. However, our syste\-matic study of higher-point correlation functions and their ensuing CFT data, precise characterization of thermal correlations, theoretical framework and concrete experimental protocol to measure boundary correlations, and simulation of an experimental setup satisfying (i)-(iv) is unprecedented. This is critical for a thorough investigation of both the duality, and consistency of the associated CFT. Here we achieve this milestone for the first time by studying interacting hyperbolic matter on the lattice.

\emph{Holographic lattice model.}---Classical or quantum theories of gravity are models where the metric tensor $g$ and distance line element $\mbox{d}s^2=\sum_{\alpha,\beta}g_{\alpha\beta}\mbox{d}x^\alpha \mbox{d}x^{\beta}$ fluctuate in time and space. In two imaginary-time spacetime dimensions, the metric can be parametrized by a single real field $\varphi(z)$ as $\mbox{d}s^2=e^{\varphi(z)}|\mbox{d}z|^2$ (isothermal coordinates), where $z=x+\rmi y$. Although the usual Einstein--Hilbert action is non-dynamical in two dimensions, a dynamical action principle for gravity results from minimizing the Liouville gravity action \cite{SeibergLG,ginsparg1993lectures}
\begin{align}
 \label{LG1} S_{\rm LG}[\varphi] = \int\mbox{d}^2z\ \Bigl[\frac{1}{2}(\nabla \vphi)^2 + \frac{2}{\ell^2}e^{\vphi}\Bigr].
\end{align}
The term $-1/\ell^2$ plays the role of a negative cosmological constant. The stationary solution $\varphi_\star(z)$ that minimizes 
the action is the hyperbolic Poincar\'{e} disk metric $\mbox{d}s_\star^2=(2\ell)^2 |\mbox{d}z|^2/(1-|z|^2)^2$. Fluctuations about this solution, 
$\varphi = \varphi_\star +\phi$, 
follow the imaginary-time action \cite{SOM}
\begin{align}
 \label{LG2} &S_{\rm grav}[\phi] = S_{\rm LG}[\varphi_\star+\phi]- S_{\rm LG}[\varphi_\star] \\
 \nonumber &= \int \frac{\mbox{d}^2z}{(1-|z|^2)^2} \Bigl[ \frac{1}{2}\phi (-\square + m^2)\phi +\frac{\phi^3}{3\ell^2}  +\frac{\phi^4}{12\ell^2}+ \dots\Bigr].
\end{align}
We may interpret $\phi(z)$ as graviton-like mode, since it parametrizes small metric fluctuations.
Importantly, dynamical gravity corresponds here to  a real scalar field $\phi$ propagating in its own hyperbolic background with $\square=(2\ell)^{-2}(1-|z|^2)^2\nabla^2$ the Laplace--Beltrami operator [ingredient (i)], and nonlinear terms like $\phi^3$ and $\phi^4$ correspond to gravitational self-interactions [ingredient (ii)].
We choose $m^2\ell^2> -1/4$ above the Breitenlohner--Freedman bound for a stable theory \cite{PhysRevLett.130.091604}.

The continuum action (\ref{LG2}) can be simulated on discrete hyperbolic lattices using the dictionary of Ref. \cite{PhysRevA.102.032208}, which yields the universal holographic lattice action
\begin{align}
 \label{circ1} S(\{\phi_\mu\}) =  -\frac{1}{2}\sum_{\mu,\nu} \phi_\mu A_{\mu\nu}\phi_\nu +\sum_{\mu} \Bigl(\frac{\hat{m}^2}{2}\phi_\mu^2+\frac{u}{3!}\phi_\mu^3\Bigr).
\end{align}
Herein, $\phi_\mu=\phi(z_\mu)$ is the field variable defined on the sites $z_\mu$ of a graph or lattice $\mathcal{G}$ with adjacency matrix $A$. 
Equation (\ref{circ1}) represents a generic tight-binding Hamiltonian realizable on the platforms discussed in the introduction, and is a universal low-energy limit of more complicated Hamiltonians. The parameters $\hat{m}^2$ and $u$ are tunable in experiment and can be matched to the Liouville action  (\ref{LG1}) \cite{SOM}. We neglect higher-order interaction terms beyond $\phi^3$ for simplicity, resulting in a model that deviates from Liouville gravity through this omission.

The choice of graph $\mathcal{G}$ determines the curved bulk space on which the action is simulated. Two-dimensional imaginary-time AdS space is emulated by hyperbolic $\{p,q\}$ lattices with $(p-2)(q-2)>4$, labelled type-I hereafter. Black holes are realized by identifying points on two geodesics in a type-I geometry. This is equivalent to a type-II ring graph \cite{chen2023adscft} obtained by a squash-and-wrap procedure, see Fig. \ref{FigBTZ}. The squash step is $z_\mu \mapsto \zeta_\mu = \frac{2}{\pi}\ln(\frac{1+z_\mu}{1-z_\mu})$,
which produces an infinite strip, from which a finite strip of width $w$ is obtained with its truncated edges identified, followed by the wrap step $\zeta_\mu \mapsto \hat{z}_\mu = e^{2\pi\rmi(\zeta_\mu+\rmi)/w}$ yielding a ring. 
When applied to the Poincar\'{e} disk metric, 
this results in a time\-slice of the three-dimensional Ba\~{n}ados--Teitelboim--Zanelli (BTZ) black hole metric \cite{banados92,carlip95,Keski1999,chen2023adscft,SOM} given by
\begin{align}
 \label{metEq} \mbox{d}s_{\rm II}^2 &= \Bigl(\frac{\ell w}{4}\Bigr)^2\frac{|\mbox{d}\hat{z}|^2}{|\hat{z}|^2\cos^2[\frac{\pi}{2}(1-\ln|\hat{z}|/\ln \hat{r}_{\rm H})]},
\end{align}
which suggests the black hole interpretation on the lattice. The black hole temperature is $T =w/8\pi\ell$ with horizon radius $\hat{r}_{\rm H}=e^{-2\pi/w}$. In our numerical construction \cite{SOM}, we have $w=kP$, where $k$ is an integer that we can choose freely and $P=P(p,q)$ is a lattice-dependent constant. By varying $k,p,q$, we can access a large, albeit discrete set of temperatures [ingredient (iii)].

\emph{Bulk-boundary-correspondence.}---To study holography, we divide the graph $\mathcal{G}$, which may be either type-I or type-II, into its bulk and boundary components, $\mathring{\mathcal{G}}$ and $\partial \mathcal{G}$, such that $\mathcal{G} = \mathring{\mathcal{G}} \cup \partial \mathcal{G}$. We label generic sites on $\mathcal{G}$ by Greek letters $\mu,\nu $, whereas bulk and boundary sites are labeled by Roman letters $i,j$ and  $a,b$, respectively. We then consider  fixed boundary field values given by $\phi_a=J_a$. The associated bulk partition function reads
\begin{align}
 \label{circ2} Z(\{J_a\}) = \int \Bigl(\prod_i\mbox{d}\phi_i\Bigr) e^{-S(\{\phi_i,\phi_a=J_a\})}.
\end{align}
Note that we only integrate over the bulk field values. The bulk-boundary-correspondence asserts that $Z(\{J_a\})$ is the generating function for a CFT on the boundary \cite{gubser98,witten1998}. The associated connected $n$-point correlation functions for some boundary field $\mathcal{O}_a$ are given by
\begin{align}
 \label{circ3} \langle \mathcal{O}_{a_1}\dots\mathcal{O}_{a_n}\rangle &= \frac{\partial^n \ln Z}{\partial J_{a_1}\cdots \partial J_{a_n}}
\Bigr|_{J=0}
\end{align}
[ingredient (iv)]. Properties like spin or scaling behavior of $\mathcal{O}_a$ are determined by the bulk theory, but no action underlying the CFT is specified by the duality \textit{per se}. An important future research inquiry is to explore whether a representative CFT action can be constructed in the holographic toy model. Furthermore, adding more fields and symmetries to the bulk action, the consistency of richer dual CFTs can be probed.

\begin{figure}[t!]
\includegraphics[width=\linewidth]{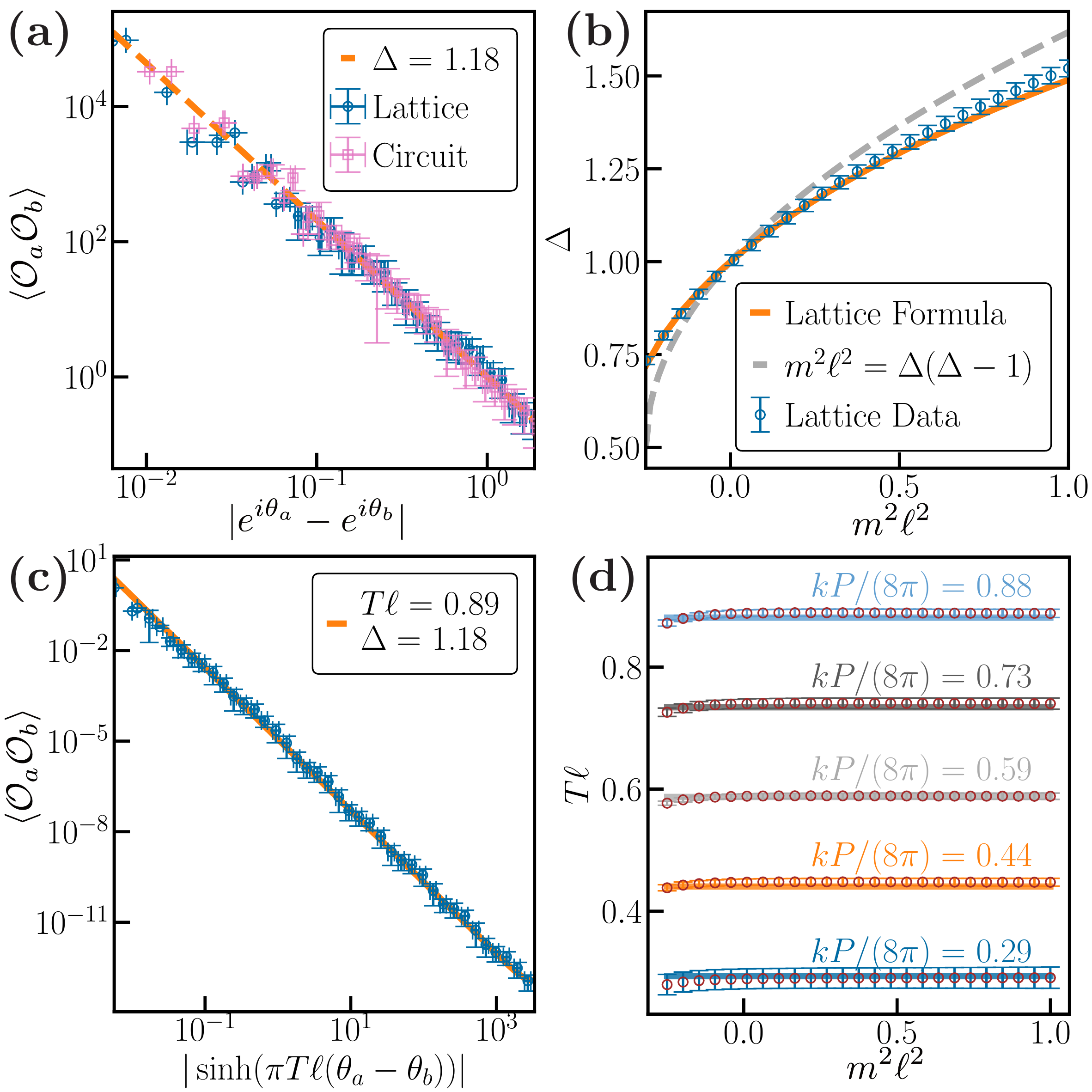}
  \caption{\textbf{CFT two-point function.} Results are shown for $\{3,7\}$ hyperbolic flakes. \textbf{(a)} Representative two-point function in type-I geometry with bulk mass $m^2\ell^2=0.271$. Data sets are computed from Eq. (\ref{circ6}) (blue) and numerical simulation of the electrical circuit in Eqs. (\ref{eom1})-(\ref{eom3}) using a realistic diode (pink). The data scatter results from binning the discrete lattice coordinates $\theta_a$.   Fitted power-law formula (\ref{CorrEq}) shown in orange. \textbf{(b)} Fitted conformal dimension $\Delta$ versus $m^2\ell^2> -1/4$ (blue squares with fit error). Due to lattice corrections captured by Eq. (\ref{fX}) (orange), $\Delta$ deviates from the continuum formula $m^2\ell^2=\Delta(\Delta-1)$ (gray dashed). \textbf{(c)} Two-point function on type-II lattice for same mass as in \textbf{a}, showing thermal behavior according to Eq. (\ref{CorrEq}). \textbf{(d)} The fitted type-II temperature $T$ (circles) is approximately independent of $m$ and described by $T\ell=kP/8\pi$ (lines). Here, $P\approx1.845$ for $\{3,7\}$ and $k=4,6,8,10,12$ in the plot.}
\label{Fig2PT}
\end{figure}

\begin{figure*}[htb!]
    \includegraphics[width=1\linewidth]{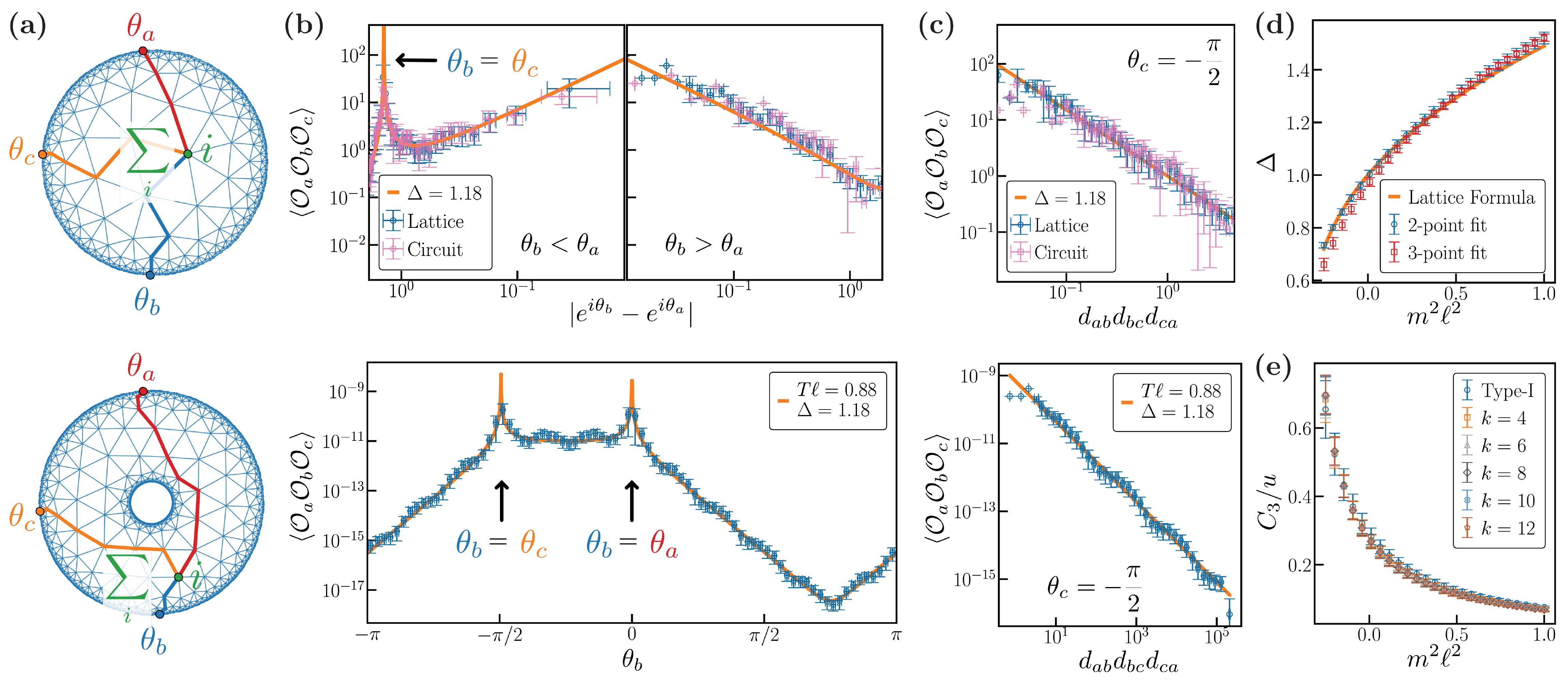}
\caption{\textbf{CFT three-point function.}  
  We visualize the three-point function on type-I (top row) and type-II geometries (bottom row). We place $\mathcal{O}_a$ at angle $\theta_a$, $\mathcal{O}_c$ at angle $\theta_c$, and vary the position of $\mathcal{O}_b$ at angle $\theta_b$. \textbf{(a)} The 3-point function corresponds to the star-shaped Witten diagram of Eq. (\ref{circ7}). The largest contribution to the sum comes from graph-geodesic paths connecting the boundary sites through an intermediate bulk site $i$. \textbf{(b)}
 Three-point functions on $\{3,7\}$ type-I and type-II flakes with parameters as in Fig. \ref{Fig2PT}. Data sets are computed from Eq. (\ref{circ7}) (blue) and numerical simulation of the electrical circuit in Eqs. (\ref{eom1})-(\ref{eom3}) using a realistic diode (pink). The solid orange lines constitute the theory prediction, where $\Delta$ and $T$ follow from the 2-point function, and only $C_3$ is a fitted parameter. Correlations diverge when any two of the angles coincide (arrows). \textbf{(c)} Scaling collapse according to the characteristic power-law behavior $\langle \mathcal{O}_a\mathcal{O}_b\mathcal{O}_c\rangle\propto (d_{ab}d_{ac}d_{bc})^{-\Delta}$  (orange), where $d_{ab}=|e^{\rmi \theta_a}-e^{\rmi \theta_b}|$ for type-I and $d_{ab}(T)=\sinh(\pi T\ell|\theta_a-\theta_b|)/\pi T \ell$ for type-II geometries, respectively.
 \textbf{(d)} Scaling dimensions $\Delta(m^2\ell^2)$ extracted from 2- and 3-point function on type-I graphs are consistent within the errors. \textbf{(e)} The normalized 3-point coefficient $C_3/u$ is determined by $m^2\ell^2$ and agrees for both type-I and type-II lattices, indicating their CFTs are equal.}
\label{Fig3PT}
\end{figure*}

The holographic properties of the lattice model (\ref{circ1}) can be studied theoretically in the perturbative regime $u\ll 1$, which mirrors the semiclassical limit of the AdS/CFT correspondence where only tree-level diagrams contribute \cite{SOM}.
The 2-point function for $u=0$ reads 
\begin{align}
 \label{circ6} \langle \mathcal{O}_a\mathcal{O}_b\rangle = - M_{ab}+\sum_{i,j}M_{ai}G_{ij}M_{jb}.
\end{align}
Here, $M_{\mu\nu}=(G^{-1})_{\mu\nu}=-A_{\mu\nu}+\hat{m}^2\delta_{\mu\nu}$ is the bare inverse propagator, with $\hat{m}^2=q+qh^2m^2\ell^2$ relating to the physical mass $m$ through a dimensionless lattice constant $h=(1-\frac{\sin^2(\pi/q)}{\cos^2(\pi/p)})^{1/2}$ \cite{PhysRevA.102.032208,petermann2023eigenmodes}. 
Approximately, $\langle \mathcal{O}_a\mathcal{O}_b\rangle \approx G_{ab}$ is the bulk propagator extrapolated to the boundary sites $a,b$.  
For the 3-point function, we have
\begin{align}
 \label{circ7}  \langle \mathcal{O}_a\mathcal{O}_b\mathcal{O}_c\rangle =u \sum_{i} B_{ai}B_{bi}B_{ci}
\end{align}
with $B_{ai}=\sum_j M_{aj}G_{ji}$ the boundary-to-bulk propagator. This is a Witten diagram, where the bulk site $i$ connects the boundary sites $a,b,c$, see Fig. \ref{Fig3PT}a, and the largest contribution comes from graph-geodesic paths.
Remarkably, this perturbative formula captures the CFT correlations also on the lattice. All $n$-point functions with $n\geq 3$ vanish for $u=0$ in our model, emphasizing again the importance of interactions embodied by $u\phi^3$ in Eq. (\ref{circ1}). A 4-point function appears at order $u^2$ \cite{ammon15}.

\emph{Correlation functions and CFT data.}---The characteristic forms of the two- and higher-point correlation functions of a CFT distinguish it from a mere scale invariant theory. In particular, the 2- and 3-point functions are fully determined by two parameters as part of their CFT data, the scaling dimension $\Delta$ and 3-point coefficient $C_3$, 
\begin{align}
 \label{CorrEq} \langle \mathcal{O}_a\mathcal{O}_b\rangle \simeq \frac{1}{(d_{ab})^{2\Delta}},\ \langle \mathcal{O}_a\mathcal{O}_b\mathcal{O}_c\rangle \simeq \frac{C_3}{(d_{ab}d_{ac}d_{bc})^\Delta}.
\end{align}
Here we normalize the boundary fields $\mathcal{O}_a$ such that the numerator of the 2-point function is unity. The function $d_{ab}$ determines the distance between the sites $a,b$ in the CFT. A hallmark of the continuum AdS/CFT correspondence is that $\Delta$ and $C_3$ are tunable by varying the bulk parameters $m^2\ell^2$ and $u$. We now show that the same is true for our holographic lattice model.

The boundary 2- and 3-point functions determined from Eqs. (\ref{circ6}) and (\ref{circ7}) are summarized in Figs. \ref{Fig2PT} and \ref{Fig3PT}, respectively. They agree with the expected behavior through the identification \cite{SOM}
\begin{align}
 d_{ab} =\begin{cases}  |e^{\rmi \theta_a}-e^{\rmi \theta_b}| & \text{(type-I)} \\ \frac{\sinh(\pi T \ell|\theta_a-\theta_b|)}{\pi T \ell} & \text{(type-II)}\end{cases}.
\end{align}
Here $\theta_{a,b}$ are the angle coordinates of the boundary sites. This emulates CFTs on a circle at zero and finite temperature. While Eqs. (\ref{CorrEq}) are good approximations on type-II lattices, especially for the 2-point function, the proper quantitative formula requires to replace $\theta_a-\theta_b\to \theta_a-\theta_b+2\pi n$ in $d_{ab}$ with a subsequent sum over $n\in\mathbb{Z}$ to make the functions periodic \cite{SOM}. This behavior, reminiscent of the method of images, also arises in the BTZ geometry and is thus expected here \cite{Keski1999}.

Our main novel findings from the analysis of correlations are that (a) the scaling dimension $\Delta(m^2\ell^2)$ extracted from the 2-point function also captures the scaling of the 3-point function, indicating a consistent CFT [Fig \ref{Fig3PT}d], (b) the parameter $k$ on type-II lattices determines the temperature consistent with the formula $T\ell= kP/8\pi$ [Fig. \ref{Fig2PT}d], but leaves the CFT data invariant, and (c) the CFT data $\Delta(m^2\ell^2)$ and $C_3(m^2\ell^2)$ for the CFTs simulated on both type-I and type-II lattices is identical [Fig. \ref{Fig3PT}d,e]. The latter finding indicates that they are the same CFTs, but at zero and finite temperature. These nontrivial results also solidify the interpretation of the type-II lattice as a geometry that emulates a black hole.

The holographic relation between $\Delta$ and $m^2\ell^2$ is of the form $m^2\ell^2=f(\Delta(\Delta-1))$ with $f_{\rm cont}(X)=X$ in the continuum, while a gradient expansion for the $\{3,7\}$-lattice yields 
\begin{align}
 \label{fX} f(X) &\simeq  X + \frac{h^2}{4}(X^2+2X)+\frac{h^4}{36}(X^3+10X^2+12X)
\end{align}
with $h=0.497$ \cite{PhysRevD.105.114503,Stegmaier2021,SOM}, see Fig. \ref{Fig2PT}b. The value of $h$ and form of $f(X)$ depend on the $\{p,q\}$ lattice, with the universal continuum limit recovered for $h\to 0$. To extract the 3-point coefficient $C_3$ in a manner that reduces scatter due to the lattice discretization, we construct the function $F_{abc} = \langle \mathcal{O}_a\mathcal{O}_b\mathcal{O}_c\rangle/[\langle\mathcal{O}_{a}\mathcal{O}_b\rangle\langle\mathcal{O}_{a}\mathcal{O}_c\rangle\langle\mathcal{O}_{b}\mathcal{O}_c\rangle]^{1/2}$ that we average over $a,b,c$ to obtain $F_{abc}\simeq C_3$ \cite{SOM}.

\emph{Experimental protocol for electrical circuits.}---We propose to realize the holographic toy model in electrical circuits by implementing the 
equation of motion
\begin{align}
 \label{eom1}-\sum_\nu A_{\mu\nu}\bar{V}_\nu + \hat{m}^2\bar{V}_\mu + \frac{u}{2}\bar{V}_\mu^2 = 0,
\end{align}
where $\phi_\mu \to \bar{V}_\mu$ is the normalized local voltage at node $z_\mu$. By applying voltage sources on the boundary, fixed values $\bar{V}_a=J_a$ can be realized. The solution $\bar{V}_\mu$ parametrically depends on the $J_a$ chosen. By applying time-dependent boundary conditions, $J_a(t)$, $t$-derivatives conveniently translate to $J_a$-derivatives below. From the measured or simulated voltage signal $\bar{V}_\mu(t)$, we compute the circuit generating function
\begin{align}
 \label{eom2} W_{\rm circ}(t) = -S(\{\bar{V}_\mu(t)\}),
\end{align}
with action $S(\{\phi_\mu\})$ from Eq. (\ref{circ1}). We have $W_{\rm circ}\simeq\ln Z$ in the saddle-point approximation, sufficient for the semiclassical holographic limit studied here.

To realize the terms in Eq. (\ref{eom1}), we use resistors to generate the linear couplings $\sum_\nu( -A_{\mu\nu} + \hat{m}^2\delta_{\mu\nu})V_\nu$, while diodes can be used for the nonlinear term $u V_\mu^2$. Indeed, the current through a diode  is approximated by the Shockley equation $I(V) = I_{\rm S}(\mathrm{e}^{V/V_{\rm S}}-1) \approx I_{\rm S}[V/V_{\rm S} + (V/V_{\rm S})^2/2]$, which yields the desired $V_\mu^2$-term after absorbing the linear part into $\hat{m}^2V_\mu$. The concrete circuit parameters are listed in Suppl. Sec. S8 \cite{SOM}. On short time scales, dissipative terms present in any realistic circuit lead to transient behavior. These effects are neglected here and the correspondence is realized in the steady state, assuming an instantaneous response to $J_a(t)$.

Our protocol to compute $2$-point functions from $W_{\rm circ}(t)$ is as follows. We apply a drive that linearly ramps $(J_a(t),J_b(t)) = (K_a,K_b)(t-t_0)$, crossing zero at $t=t_0$, while all other $J_{\mu}(t)\equiv 0$ for $\mu\neq a,b$. Since all second time-derivatives vanish, we have
\begin{align}
 \label{eom3} \frac{\mbox{d}^2W_{\rm circ}}{\mbox{d}t^2}\Bigr|_{t=t_0} &= \sum_{\mu,\nu} \dot{J}_\mu \dot{J}_\nu \frac{\partial^2W_{\rm circ}}{\partial J_\mu\partial J_\nu}\Bigr|_{J=0}\\
 \nonumber &=K_a^2 \langle\mathcal{O}_a^2\rangle + 2K_aK_b \langle\mathcal{O}_a\mathcal{O}_b\rangle + K_b^2 \langle \mathcal{O}_b^2\rangle.
\end{align}
By choosing three linearly independent ramps in a series of measurements, e.g. $(K_a,K_b)=(1,0),(0,1),(1,1)$, this set of three linear equations can be solved for $\langle \mathcal{O}_a\mathcal{O}_b\rangle$. Similarly, to measure $n$-point functions, we ramp $n$ boundary sites $J_{a_1}(t),\dots,J_{a_n}(t)$ linearly. By using ten linearly independent ramps, we obtain  $\langle \mathcal{O}_a\mathcal{O}_b\mathcal{O}_c\rangle$ from $\mbox{d}^3W_{\rm circ}/\mbox{d}t^3|_{t=t_0}$. 

In Fig. \ref{Fig2PT}a we present results for the so-obtained 2-point function using an LTspice simulation of an electrical circuit with a realistic Schottky diode (model RBE1VAM20A). In Fig. \ref{Fig3PT}b,c we present the 3-point function for the same parameters from a Mathematica simulation of Eqs. (\ref{eom1})-(\ref{eom3}). Since the signals in actual electrical circuits are known to be close to such numerical simulations \cite{PhysRevLett.114.173902,PhysRevX.5.021031,lee2018topolectrical}, this serves as a proof of principle that our protocol is experimentally feasible.

\emph{Acknowledgments.}---We thank Joseph Maciejko for profound contributions especially in the early stages of the project and his insightful comments on the manuscript. We are grateful for fruitful discussions with Giuseppe Di Giulio, Dominik Neuenfeld, and Canon Sun.
AC was supported by the Natural Sciences and Engineering Research Council of Canada (NSERC) Discovery Grant RGPIN-2020-06999, the Avadh Bhatia Fellowship, and the Faculty of Science at the University of Alberta. AC and IB acknowledge support through the University of Alberta startup fund UOFAB Startup Boettcher. 
SD was supported by the Faculty of Science at the University of Alberta. 
MK was supported, in part, by the U.S.~Department of Energy grant DE-SC0012447. 
 IB acknowledges funding from the NSERC Discovery Grants RGPIN-2021-02534 and DGECR2021-00043. PB, AF, MG, PML, RM, RS, AS, RT, and JE acknowledge support by Germany's Excellence Strategy through the W\"urzburg‐Dresden Cluster of Excellence on Complexity and Topology in Quantum Matter ‐ ct.qmat (EXC 2147, project‐id 390858490), and by the Deutsche Forschungsgemeinschaft (DFG) through the Collaborative Research Center ``ToCoTronics”, Project-ID 258499086—SFB 1170.
PML was supported by the Ambizione grant No.~185806 by the Swiss National Science Foundation (SNSF) and the European Union (ERC, QuSimCtrl, 101113633). Views and opinions expressed are however those of the authors only and do not necessarily reflect those of the European Union or the European Research Council Executive Agency. Neither the European Union nor the granting authority can be held responsible for them.

SD and AC contributed equally to this work.

\bibliography{refs_holo}

\clearpage

\cleardoublepage

\setcounter{equation}{0}
\renewcommand{\theequation}{S\arabic{equation}}
\setcounter{figure}{0}
\renewcommand{\thefigure}{S\arabic{figure}}

\begin{center}
\textbf{\Large Supplemental Material}
\end{center}

\noindent This supplemental material details the algorithm for constructing hyperbolic lattice flakes, the reason for interpreting type-II as BTZ geometry, our fit procedure, the derivation of Eqs. (2), (7), (8), (10), and (11), and the electrical circuit simulation.

\tableofcontents

\section{S1. Derivation of Eq. (2)}\label{SeqLiou}

We write the metric in isothermal coordinates as
\begin{align}
 \mbox{d}s^2 = g_{ij} \mbox{d}x^i\mbox{d}x^j = e^{\varphi} (\mbox{d}x^2+\mbox{d}y^2)
\end{align}
with $g_{ij} = e^{\varphi} \delta_{ij}$ and $i,j=1,2$. We have $g^{ij}=e^{-\varphi}\delta_{ij}$ and $\sqrt{\mbox{det}(g_{ij})} = e^{\varphi}$. Define the Gaussian curvature
\begin{align}
 \mathcal{K}[g] = -\frac{1}{2}e^{-\varphi}\nabla^2\varphi.
\end{align}
The Poincar\'{e} disk metric
\begin{align}
\mbox{d}s_\star^2 &= (2\ell)^2 \frac{\mbox{d}x^2+\mbox{d}y^2}{(1-|z|^2)^2},\\
\varphi_\star &= \ln\Bigl(\frac{(2\ell)^2}{(1-x^2-y^2)^2}\Bigr)
\end{align}
satisfies $\mathcal{K}[g_\star]=-\frac{1}{\ell^2}$.

The equations of motion associated to the Liouville action in Eq. (1) are
\begin{align}
 0 = \frac{\delta S_{\rm LG}}{\delta \vphi}[\vphi_\star] = -\nabla^2\vphi_\star +\frac{2}{\ell^2} e^{\vphi_\star},
\end{align}
which is equivalent to $\mathcal{K}[g_\star]=-\frac{1}{\ell^2}$, hence solved by the hyperbolic metric. We write $S_{\rm LG}=S_{\rm kin}+S_{\rm pot}$. Expanding $\vphi = \vphi_\star +\phi$ we obtain for the kinetic part
\begin{align}
 \nonumber &S_{\rm kin}[\vphi_\star+\phi] = \int\mbox{d}^2z\Bigl[ \frac{1}{2}[\nabla(\vphi_\star+\phi)]^2\Bigr]\\
 \nonumber &= \int\mbox{d}^2z\Bigl[ \frac{1}{2}(\nabla\vphi_\star)^2+\nabla\vphi_\star\cdot\nabla\phi+\frac{1}{2}(\nabla\phi)^2\Bigr]\\
  \nonumber &= \int\mbox{d}^2z\Bigl[ -\frac{1}{2}\vphi_\star\nabla^2\vphi_\star-\phi\nabla^2\vphi_\star-\frac{1}{2}\phi\nabla^2\phi\Bigr]\\
 \nonumber &= \int\mbox{d}^2z\ e^{\vphi_\star} \Bigl[ -\frac{1}{2}\vphi_\star \frac{2}{\ell^2}-\phi\frac{2}{\ell^2}-\frac{1}{2}\phi(e^{-\vphi_\star}\nabla^2)\phi\Bigr].
\end{align}
For the potential part we obtain
\begin{align}
\nonumber S_{\rm pot}[\vphi_\star+\phi] &= \int\mbox{d}^2z \Bigl[ \frac{2}{\ell^2}e^{\vphi_\star}e^\phi\Bigr]=\int\mbox{d}^2z\ e^{\vphi_\star} \Bigl[ \frac{2}{\ell^2}e^\phi\Bigr].
\end{align}
Adding both expansions and neglecting terms that are independent of $\phi$ we obtain
\begin{align}
  S_{\rm grav}[\phi] = \int\mbox{d}^2z\ e^{\vphi_\star}\Bigl[ -\frac{1}{2}\phi(e^{-\vphi_\star}\nabla^2)\phi+\frac{2}{\ell^2}(e^\phi-\phi)\Bigr].
\end{align}
Note that this action is again of Liouville form \cite{ginsparg1993lectures}, but for a hyperbolic fiducial metric, whereas the starting action $S_{\rm LG}$ was based on a flat fiducial metric. This reflects the shift invariance of the Liouville action. However, due to the saddle-point expansion around $\vphi_\star$, fluctuations parametrized by $\phi$ are expected to be small, whereas this is not true for the field $\vphi$. Expanding the potential term in powers of $\phi$ we arrive at
\begin{align}
 \nonumber  S_{\rm grav}[\phi] ={}& \int\mbox{d}^2z\ e^{\vphi_\star}\Bigl[ -\frac{1}{2}\phi(e^{-\vphi_\star}\nabla^2)\phi\\
  &+\frac{2}{\ell^2}\Bigl(\frac{1}{2}\phi^2+\frac{1}{3!}\phi^3+\frac{1}{4!}\phi^4+\dots\Bigr)\Bigr]\\
 \nonumber ={}&\int\frac{\mbox{d}^2z\ (2\ell)^2}{(1-|z|^2)^2}\Bigl[\frac{1}{2}\phi\Bigl(-\square+\frac{2}{\ell^2}\Bigr)\phi\\
 &+ \frac{1}{3\ell^2}\phi^3+\frac{1}{12\ell^2}\phi^4+\dots\Bigr].
\end{align}
This is of the form of Eq. (2) after appropriate choice of constants $m,u,\mu$, and normalization of $\phi$. Here we used
\begin{align}
 e^{\vphi_\star} &=\frac{(2\ell)^2}{(1-|z|^2)^2},\\
 \square&= e^{-\vphi_\star}\nabla^2 = \frac{1}{(2\ell)^2}(1-|z|^2)^2\nabla^2.
\end{align}

To connect the Liouville action to the lattice action
\begin{align}
 S(\{\phi_\mu\}) =  -\frac{1}{2}\sum_{\mu,\nu} \phi_\mu A_{\mu\nu}\phi_\nu +\sum_{\mu} \Bigl(\frac{\hat{m}^2}{2}\phi_\mu^2+\frac{u}{3!}\phi_\mu^3\Bigr),
\end{align}
we replace $\sum_\nu A_{\mu\nu}\phi_\nu \to (q +q h^2\ell^2\square )\phi_\mu$ \cite{PhysRevA.102.032208} and find
\begin{align}
\nonumber S(\{\phi_\mu\}) &=  \sum_{\mu} \Bigl(\frac{1}{2}\phi_\mu(-qh^2\ell^2\square-q +\hat{m}^2 )\phi_\mu+\frac{u}{3!}\phi_\mu^3\Bigr)\\
 &=  \sum_{\mu} \Bigl(qh^2\ell^2\frac{1}{2}\phi_\mu(-\square+m^2)\phi_\mu+\frac{u}{3!}\phi_\mu^3\Bigr).
\end{align}
We inserted $\hat{m}^2=q+qh^2\ell^2m^2$. Through this equation, the experimental parameters $\hat{m}^2$ and $u$ can be connected to the continuum Liouville action S$_{\rm grav}[\phi]$.

\section{S2. Construction of type-I and II flakes}\label{SecFlake}

To construct hyperbolic flakes, we start from regular hyperbolic \{$p,q$\} tilings with $(p-2)(q-2)>4$, tessellated by $p$-sided polygons with $q$ of them meeting at each site. We refer to type-I flakes as hyperbolic lattices with open boundary conditions, generated via vertex inflation \cite{PhysRevX.10.011009,Jahn2020,Chen2023Symmetry} and embedded in the Poincar\'{e} disk
\begin{align}
\mathbb{D}&=\left\{ z\in\mathbb{C},\ |z|<1\right\},\\
\mbox{d}s^2 &= (2\ell)^2 \frac{|\mbox{d}z|^2}{(1-|z|^2)^2}.
\end{align}
Type-II flakes are constructed from the type-I flakes by identifying two geodesics related by a boost (as illustrated in Fig.~\ref{FigBTZ}), akin to the construction of BTZ black holes from the AdS space. While such identification can be done by modifying the adjacency matrix of the lattice without a change of site coordinates, as the adjacency matrix alone encodes the lattice connectivity and thereby dictates the subsequent computation of correlation functions, we perform two conformal transformations of the site coordinates to facilitate the construction of finite-sized type-II flakes and clear visualization of their topologically nontrivial geometry.

\begin{figure*}[t!]
    \includegraphics[width=1\linewidth]{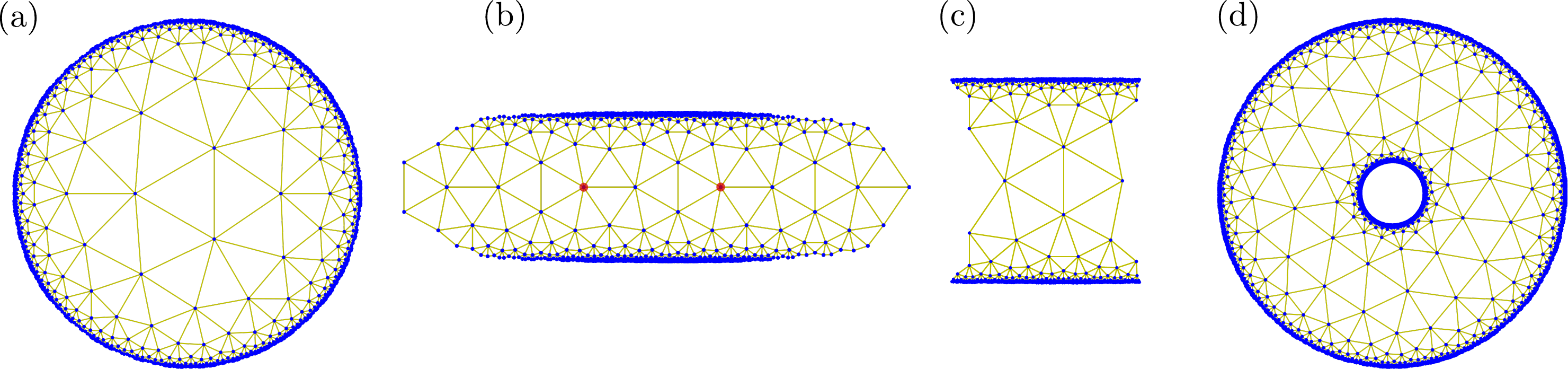}
\caption{\textbf{Construction of type-II hyperbolic lattices.} (a) A type-I hyperbolic lattice with \{$3,7$\} tessellation generated via finite iterations of vertex inflation. (b) After the conformal transformation in Eq.~\eqref{eq:C1}, the finite-sized lattice is "squashed" into a finite strip with a discrete translation symmetry along the $x$-direction. The translation symmetry is incomplete due to the finite system size. The period $P$ is given by the distance between the highlighted sites in red. (c) We isolate the central periodic segment in the range $x\in[-P/2,P/2)$ (shown here), where the CFT boundaries are the most complete, and join together $k$ copies of the segment to form a finite periodic strip of width $w=kP$. (d) After joining the ends of the strip and passing the coordinates through a second conformal map defined in Eq.~\eqref{eq:C2}, we arrive at a type-II hyperbolic lattice that represents the discretized time-slice of a BTZ black hole with temperature $T=kP/8\pi$.
 }
\label{3-7_construction}
\end{figure*}

In our analysis, we focus on the $\{3,7\}$ tiling because using triangles as tiles allows every boundary site to be connected to a bulk site, leading to ample data for the correlation functions. We generate a finite-sized type-I $\{3,7\}$ lattice, as depicted in Fig.~\ref{3-7_construction}a, via the vertex-inflation algorithm detailed in Ref.~\onlinecite{Chen2023Symmetry}. Specifically, we implement six or seven iterations of vertex inflation (dubbed "shells"), resulting in a total of $N=960$ or 2523 sites, respectively. We denote the adjacency matrix of the type-I lattice as $A$ and the Poincar\'{e} disk coordinates as $\{z_{\mu}\}_{\mu=1}^{N}$.

We then proceed to construct type-II hyperbolic lattices from the 6-shell $\{3,7\}$ type-I lattice. We first apply a conformal transformation $\mathcal{C}_{1}$ from the Poincar\'{e} disk to strip coordinates:
\begin{equation} 
\mathcal{C}_{1}:z\mapsto \zeta=\frac{2}{\pi}\ln\left(\frac{1+z}{1-z}\right), \label{eq:C1}
\end{equation} 
with domain and range 
\begin{equation} 
\mathcal{C}_{1}:\mathbb{D}\rightarrow\mathbb{B}=\left\{\zeta\in \mathbb{C},|\text{Im}(\zeta)|<1\right\}. 
\end{equation} 
$\mathcal{C}_{1}$ maps $0$ to $0$ and the upper and lower halves of the Poincar\'{e}-disk boundary $z=e^{\text{i}\phi}$ to the lines $\text{i}+\mathbb{R}$ and $-\text{i}+\mathbb{R}$, respectively. The metric of the strip model is 
\begin{equation} 
\text{d}s^{2}=\ell^{2}\left(\frac{\pi}{2}\right)^{2}\frac{|\text{d}\zeta|^2}{\cos^{2}\left(\frac{\pi}{2}\text{Im(}\zeta)\right)}, 
\end{equation} 
which diverges at the asymptotic boundary at $\text{Im}(\zeta)\rightarrow\pm1$. The strip model is periodic along the $x$-direction with the period dependent on $p$ and $q$. Given that our type-I lattice is finite-sized, the resultant strip model  with site coordinates $\{\zeta_{\mu}\}_{\mu=1}^{N}$ (Fig.~\ref{3-7_construction}b) is non-uniform with progressively less complete asymptotic boundary away from $\text{Re}(\zeta)=0$. Upon closer inspection, we identify two sites (highlighted in red in Fig.~\ref{3-7_construction}b) which lie on the left and right vertical edges that define a periodic segment. The difference of their coordinates gives the period $P \approx 1.845$ for the $\{3,7\}$ tiling. 

Our strategy is to create a mostly homogeneous type-II flake is to take the central periodic segment between the $x$-range $[-P/2,P/2)$, where the  asymptotic boundaries at $\pm\text{i}+\mathbb{R}$ are the most complete, and copy this small segment $k$ times, joining them together to form a long strip of length $kP$. To achieve this, we single out those sites in Fig.~\ref{3-7_construction}b that lie outside the $x$-range $[-P/2,P/2)$ and truncate them from the adjacency matrix $A$ by removing the corresponding rows and columns. We also remove these sites from the list of all coordinates $\{\zeta_{\mu}\}_{\mu=1}^{N}$. The truncated segment, shown in Fig.~\ref{3-7_construction}c, has $N'=686$ sites; its adjacency matrix is denoted by $a$. Next, we begin building the adjacency matrix of the long strip, denoted $\tilde{A}$, by computing the Kronecker product of the $k\times k$ identity matrix and $a$: $\tilde{A}=\mathbb{1}_{k\times k}\otimes a$. The coordinates of the long strip are copies of the small-segment coordinates, shifted horizontally by integer multiples of $P$: $0,\ P,\ 2P,\ \dots,\ (k-1)P$. 

At this stage, our strip is formed from $k$ disjoint segments. While the coordinates are correct, the adjacency matrix does not include edges connecting the neighboring segments to maintain a coordination number of 7 on every site. To connect the segments, we need to first determine the edges that were severed at the truncation step. The procedure is as follows. For each site $\mu$ in the truncated segment (Fig.~\ref{3-7_construction}c), we identify its nearest neighbors in the untruncated band (Fig.~\ref{3-7_construction}b) using the original adjacency matrix $A$. If a particular nearest neighbor $\nu$ is not included in the truncated segment, we numerically search for the site $\nu'$ such that it belongs to the segment and equivalent to $\nu$ modulo horizontal translation defined by the periodicity. Note that during this identification process, we set the numerical tolerance in distance comparison to $10^{-10}$. We record all the severed edges on the right side of the truncated segment in a list ${(\mu,\nu')}$. We also check that the severed edges on the left side correspond to the same set of edges. Finally, we re-introduce these severed edges to the adjacency matrix $\tilde{A}$ between every pair of neighboring segments centered at $cP$ and $(c+1)P$ for $c=0,...,k-2$, as well as the pair at the ends centered at $(k-1)P$ and 0 to respect the periodic boundary condition. At this stage, $\tilde{A}$ correctly describes a type-II lattice with $kN'$ sites.

We apply another conformal transformation $\mathcal{C}_{2}$ to wrap the strip model into an annulus geometry (Fig.~\ref{3-7_construction}d): 
\begin{equation} 
\mathcal{C}_{2}:\zeta\mapsto \hat{z}=e^{2\pi\text{i}(\zeta+\text{i})/(kP)}  \label{eq:C2},
\end{equation} 
with domain and range 
\begin{equation} 
\mathcal{C}_{2}:\mathbb{B}\rightarrow\mathbb{A}=\{\hat{z}\in\mathbb{C},\ \hat{r}_{0}<|\hat{z}|<1\}. 
\end{equation} 
$\mathcal{C}_{2}$ maps the upper boundary $\text{i}+\mathbb{R}$ in the strip model to the inner boundary of the annulus at radius $\hat{r}_{0}=\mathcal{C}_2(\rmi)= e^{-4\pi/(kP)}$ and the lower boundary $-\text{i}+\mathbb{R}$ to the outer boundary at $1$. The new metric is 
\begin{align} 
 \label{EqS18} \text{d}s^{2}&=\left(\frac{\ell kP}{4}\right)^{2}\frac{|\text{d}\hat{z}|^2}{|\hat{z}|^{2}\cos^{2}\left(\frac{\pi}{2}(1+\frac{kP}{2\pi}\log|\hat{z}|)\right)}\\
 \nonumber &=\left(\frac{\ell kP}{4}\right)^{2}\frac{|\text{d}\hat{z}|^2}{|\hat{z}|^{2}\sin^{2}\left(\frac{kP}{4}\log|\hat{z}|\right)}, 
\end{align} 
which diverges at the boundaries $|\hat{z}|=\hat{r}_{0}$ and 1. The circle separating the inner and outer halves of the type-II lattice is at radius 
\begin{equation} 
\hat{r}_{H}=\mathcal{C}_{2}(0)=e^{-2\pi/(kP)}=\sqrt{\hat{r}_{0}},
\end{equation} 
which can be interpreted as the black hole horizon.

\section{S3. Relation between type-II and BTZ geometry}\label{SecBTZ}

In this section, we show that timeslices of the three-dimensional BTZ black hole correspond to the type-II geometry simulated on our hyperbolic lattices. For this purpose, we first recall some facts about the BTZ solution following Ref. \cite{carlip95}, and then show explicitly how timeslices are embedded in the Poincar\'{e} upper half-plane and disk models, $\mathbb{H}$ and $\mathbb{D}$, respectively.

We consider the imaginary-time BTZ black hole, which is a solution to the imaginary-time Einstein equations in vacuum in $2+1$ spacetime dimensions for a negative cosmological constant $\Lambda=-1/\ell^2$, given by the metric
\beq
\mbox{d}s_{\rm BTZ}^2=\left(\frac{\rho^2}{\ell^2}-M\right)\mbox{d}\tau^2+\left(\frac{\rho^2}{\ell^2}-M\right)^{-1}\mbox{d}\rho^2 +\rho^2\mbox{d}\vartheta^2.
  \label{metricEBTZ}
\eeq
Here $M$ is a dimensionless parameter and $\rho \geq\rho_{\rm H}$ is a non-compact radial coordinate that describes the region outside the event horizon, which is located at the singular radius 
\begin{align}
\rho_{\rm H}=\sqrt{M}\ell.
\end{align}
The imaginary time and radial coordinates, $\tau$ and $\vartheta$, are compactified to the intervals $\tau\in[0,2\pi\ell^2/\rho_{\rm H})$ and $\vartheta\in[0,2\pi)$ through the identifications
\begin{align}
  \label{btz_id1} (\rho,\tau,\vartheta)&\sim \Bigl(\rho,\tau+2\pi\frac{\ell^2}{\rho_{\rm H}},\vartheta\Bigr),\\
  \label{btz_id2} (\rho,\tau,\vartheta)&\sim (\rho,\tau,\vartheta+2\pi).
\end{align}
The periodicity in $\tau$ with period $T^{-1}$ corresponds to a black hole temperature
\begin{align}
 \label{EqT} T = \frac{\rho_{\rm H}}{2\pi \ell^2} =\frac{\sqrt{M}}{2\pi \ell}.
\end{align}

As any solution to the vacuum Einstein field equations in 2+1 dimensions with negative cosmological constant, the BTZ black hole is equivalent to a quotient $\mathcal{H}_3/{\sim}$, where  $\mathcal{H}_3$ is hyperbolic three-space with constant negative curvature, and $\sim$ is an equivalence relation. Here we denote $\mathcal{H}_3$ in the upper half-space coordinates
\begin{align}
 \mathcal{H}_3 &= \{(x,y,z)\in\mathbb{R}^3,\ z>0\},\\
 \mbox{d}s^2 &= \frac{\ell^2}{z^2}(\mbox{d}x^2+\mbox{d}y^2+\mbox{d}z^2).
\end{align}
The equivalence is made explicit by the coordinate transformation
\begin{align}
\nonumber x&=\Bigl(1-\frac{\rho_{\rm H}^2}{\rho^2}\Bigr)^{1/2}\cos\Bigl(\frac{\rho_{\rm H}\tau}{\ell^2}\Bigr)e^{\sqrt{M}\vartheta},\\
\label{coordinate_transform}y&=\Bigl(1-\frac{\rho_{\rm H}^2}{\rho^2}\Bigr)^{1/2}\sin\Bigl(\frac{\rho_{\rm H}\tau}{\ell^2}\Bigr)e^{\sqrt{M}\vartheta},\\
\nonumber z&=\frac{\rho_{\rm H}}{\rho}e^{\sqrt{M}\vartheta}.
\end{align}
The identification in Eqs. (\ref{btz_id1}) and (\ref{btz_id2}) limits the image to a region $\mathcal{S}_3/{\sim}$ given by
\begin{align}
 \mathcal{S}_3=\Bigl\{ (x,y,z)\in\mathcal{H}_3,\ 1 \leq \sqrt{x^2+y^2+z^2} < e^{2\pi \sqrt{M}}\Bigr\}
\end{align}
with the identification
\begin{align}
(x,y,z) \sim e^{2\pi\sqrt{M}}(x,y,z).
\end{align}
This consists of the region between two hemispheres in the upper half-space, where the inner and outer hemisphere are identified such that identified points lie on rays in $\mathcal{H}_3$ that go through the origin $(0,0,0)$. Timeslices of the BTZ geometry with a fixed value of $\tau$ correspond to $\mbox{d}\tau=0$ in Eq. (\ref{metricEBTZ}), They are described by the static black hole metric 
\begin{align}
 \label{BTZstatic} \mbox{d}s^2=\left(\frac{\rho^2}{\ell^2}-M\right)^{-1}\mbox{d}\rho^2 +\rho^2\mbox{d}\vartheta^2.
\end{align}

\begin{figure}
    \includegraphics[width=\linewidth]{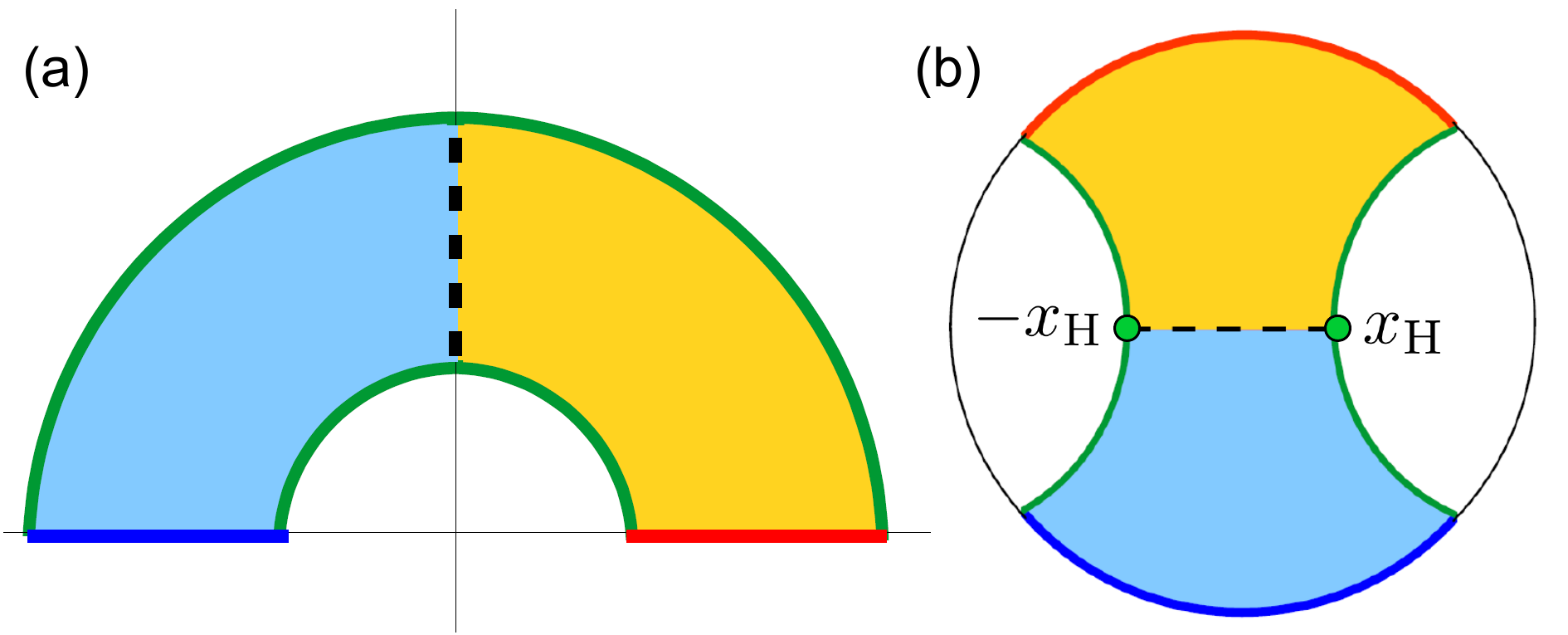}
  \caption{Timeslice of the BTZ black hole with $\tau=0$ after mapping to the upper half-plane model $\mathbb{H}$ (left) and Poincar\'{e} disk model $\mathbb{D}$ (right). Points along the geodesics indicated in green are identified; this reveals the quotient structure of the geometry. We show the points $\pm x_{\rm H}$ in $\mathbb{D}$ that are mapped to $\pm w/2$ in the strip coordinates under $\mathcal{C}_2$, which gives the relation $T=w/(8\pi \ell)$.}
  \label{FigBTZid}
\end{figure}

\emph{Embedding timeslices into $\mathbb{H}$.} We define the two-dimensional Poincar\'{e} upper half-plane model through
\begin{align}
  \mathbb{H} &= \{ x+\rmi z\in\mathbb{C},\ z>0\},\\
 \mbox{d}s^2 &= \frac{\ell^2}{z^2}(\mbox{d}x^2+\mbox{d}z^2).
\end{align}
Choosing $\tau_+=0$ or $\tau_-=\frac{1}{2}T^{-1}=\frac{\pi\ell^2}{\rho_{\rm H}}$ in Eqs. (\ref{coordinate_transform}) yields
\begin{align}
\nonumber x&=\pm \Bigl(1-\frac{\rho_{\rm H}^2}{\rho^2}\Bigr)^{1/2}e^{\sqrt{M}\vartheta},\\
\label{Hmap} y&=0,\\
\nonumber z&=\frac{\rho_{\rm H}}{\rho}e^{\sqrt{M}\vartheta},
\end{align}
where the plus (minus) sign in $x$ corresponds to $\tau_+$ ($\tau_-$). Discarding the $y$-variable, this yields Universes I and II embedded into $\mathbb{H}$ given by $\mathcal{S}_{2,\rm I}/{\sim}$ and $\mathcal{S}_{2,\rm II}/{\sim}$, respectively, with
\begin{align}
 \mathcal{S}_{2,\rm I} &= \{(x,z)\in\mathbb{H},\ 1\leq \sqrt{x^2+z^2} < e^{2\pi \sqrt{M}},\ x>0\},\\
 \mathcal{S}_{2,\rm II} &= \{(x,z)\in\mathbb{H},\ 1\leq \sqrt{x^2+z^2} < e^{2\pi \sqrt{M}},\ x<0\}
\end{align}
and the identification
\begin{align}
 (x,z) \sim e^{2\pi\sqrt{M}}(x,z).
\end{align}
The event horizon ($\rho=\rho_{\rm H}$) corresponds to the topological circle $\mathcal{S}_{2,\rm H}/{\sim}$ with
\begin{align}
\mathcal{S}_{2,\rm H} = \{(0,z)\in \mathbb{H},\ 1\leq z<e^{2\pi\sqrt{M}}\}.
\end{align}
This corresponds to a region between two geodesics (circular arcs) that are identified. See Fig. \ref{FigBTZid}a for a visualization.

\emph{Embedding timeslices into $\mathbb{D}$.} We achieved to show that timeslices $\tau_{\pm}$ corresponds to the region between two geodesics that are identified. This is the starting point of the squash-and-wrap procedure and thus proves the equivalence to the type-II geometry. To make the correspondence even clearer, we formulate the timeslices in the Poincar\'{e} disk geometry. The mapping between $\mathbb{H}$ and $\mathbb{D}$ is facilitated by the Cayley transformation. 
\begin{align}
 \mathbb{H}\ &\to\ \mathbb{D},\\
 w\ &\mapsto\ z = \frac{w-\rmi}{w+\rmi}.
\end{align}
Here we work with the complex conjugate of the Cayley map for convenience and choose $\vartheta\in[-\pi,\pi)$ to obtain a more symmetric image. The mapping in Eqs. (\ref{Hmap}) translate to Poincar\'{e} disk coordinates $z=x+\rmi y\in\mathbb{D}$ with
\begin{align}
 \label{BTZd1} x &= \frac{\frac{\rho}{\rho_{\rm H}}\sinh(\sqrt{M}\vartheta)}{1+\frac{\rho}{\rho_{\rm H}}\cosh(\sqrt{M}\vartheta)},\\
 \label{BTZd2} y &= \frac{\pm 1\sqrt{\frac{\rho^2}{\rho_{\rm H}^2}-1}}{1+\frac{\rho}{\rho_{\rm H}}\cosh(\sqrt{M}\vartheta)}.
\end{align}
The static BTZ metric in Eq. (\ref{BTZstatic}) is mapped to the hyperbolic metric
\begin{align}
 \mbox{d}s^2 = (2\ell)^2 \frac{|\mbox{d}z|^2}{(1-|z|^2)^2}.
\end{align}
The image splits into Universes I and II in $\mathbb{D}$ with points along a geodesic identified, see Figs. 1b and \ref{FigBTZid}b. Along the horizon with $\rho=\rho_{\rm H}$ we have $y=0$ and $-x_{\rm H}\leq x<x_{\rm H}$ with opposite ends identified, and 
\begin{align}
x_{\rm H} = \tanh\Bigl(\frac{\rho_{\rm H}\vartheta_{\rm max}}{2\ell}\Bigr) = \tanh\Bigl(\frac{\rho_{\rm H}\pi}{2\ell}\Bigr).
\end{align}
Under the map $\mathcal{C}_2$ from Eq. (\ref{eq:C2}), $x_{\rm H}$ is mapped to the point $\frac{w}{2}$ in the strip geometry, and hence
\begin{align}
\frac{w}{2} = \mathcal{C}_2(x_{\rm H}) = \frac{2}{\pi}\ln\Bigl(\frac{1+x_{\rm H}}{1-x_{\rm H}}\Bigr) = \frac{4}{\pi} \frac{\rho_{\rm H}\pi}{2\ell} = \frac{2\rho_{\rm H}}{\ell}.
\end{align}
Using Eq. (\ref{EqT}), we conclude that the relation between $T$ and $w$ in the type-II geometries is given by
\begin{align}
 \label{EqS45} T = \frac{w}{8\pi \ell}.
\end{align}

\section{S4. Derivation of Eqs. (7) and (8)}\label{SecCorr}

We divide the graph $\mathcal{G}$ into bulk (interior) and boundary and use the following notation:
\begin{center}
\begin{tabular}{|c|c|}
\hline
 graph & \ site index \ \\
\hline\hline
 \ $\mathcal{G}=\mathring{\mathcal{G}}\cup \partial\mathcal{G}$ \ & $\mu, \nu, \dots$ \\
\hline
 $\mathring{\mathcal{G}}$ & $i,j,\dots$  \\
\hline
 $\partial \mathcal{G}$ & $,b,\dots$  \\
\hline
\end{tabular}
\end{center}
The quadratic part of the action is
\begin{align}
 S_0(\{\phi_\mu\}) = \frac{1}{2}\sum_{\mu,\nu}M_{\mu\nu} \phi_\mu\phi_\nu
\end{align}
with $M=-A+\omega\mathbb{1}$ a real symmetric matrix. We set the values of $\phi_a$ on $\partial \mathcal{G}$ to $J_a$ and consider the generating function
\begin{align}
\nonumber Z_0(\{J_a\}) &= \int \Bigl(\prod_i\mbox{d}\phi_i\Bigr) e^{-S_0(\{\phi_i,\phi_a=J_a\})}\\
&=e^{-S_{0,\rm eff}(\{J_a\})}
\end{align}
with
\begin{align}
S_{0,\rm eff} = \frac{1}{2}\sum_{a,b} \Bigl(M_{ab}-\sum_{i,j}M_{ai}G_{ij}M_{jb}\Bigr)J_aJ_b
\end{align}
and $G_{ij}=(M^{-1})_{ij}$. 

\emph{Boundary two-point function.} We obtain the non-interacting two-point boundary correlator
\begin{align}
\nonumber \langle \mathcal{O}_a\mathcal{O}_b\rangle_{0}  &= \frac{\partial^2\ln Z_0}{\partial J_a\partial J_b} = -\frac{\partial^2S_{\rm eff}}{\partial J_a\partial J_b}\\
&= -M_{ab} + \sum_{i,j}M_{ai}(M^{-1})_{ij}M_{jb}.
\end{align}
For $u=0$ this coincides with $\langle \mathcal{O}_a\mathcal{O}_b\rangle$ and yields Eq. (7). We neglect here the subleading self-energy term that arises for $u\neq 0$, as is commonly done in holographic studies.

\emph{Boundary three-point function.} Note that the non-interacting three-point boundary correlator vanishes,
\begin{align}
\langle \mathcal{O}_a\mathcal{O}_b\mathcal{O}_c\rangle_{0} = \frac{\partial^3\ln Z_0}{\partial J_a\partial J_b\partial J_c}=0,
\end{align}
as do higher $n$-point functions. Including the cubic interaction term we have 
\begin{align}
S(\{\phi_\mu\}) = S_0(\{\phi_\mu\})+\frac{u}{3!}\sum_\mu  \phi_\mu^3.
\end{align}
We treat interactions perturbatively according to
\begin{align}
 \nonumber &Z(\{J_a\}) \\
 \nonumber &=e^{-\frac{u}{3!}\sum_a J_a^3} \int\Bigl(\prod_i \mbox{d}\phi_i\Bigr) e^{-S_0(\{\phi_i,J_a\})-\frac{u}{3!}\sum_i \phi_i^3}\\
 \nonumber  &\simeq e^{-\frac{u}{3!}\sum_a J_a^3} \int\Bigl(\prod_i \mbox{d}\phi_i\Bigr)\Bigl(1-\frac{u}{3!}\sum_j \phi_j^3\Bigr) e^{-S_0(\{\phi_i,J_a\})}\\
  &=e^{-\frac{u}{3!}\sum_a J_a^3}Z_0(\{J_a\})\Bigl(1- \frac{u}{3!} \sum_i\langle \phi_i^3\rangle_0\Bigr).
\end{align}
We have
\begin{align}
\langle \phi_i^3\rangle_0 &=3\langle\phi_i^2\rangle_{0,\rm con} \langle \phi_i\rangle_0 +\langle \phi_i\rangle_0\langle \phi_i\rangle_0\langle \phi_i\rangle_0\\
&=:\sum_a J_a w_a^{(i)}+ \sum_{a,b,c} J_aJ_bJ_c w_{abc}^{(i)}
\end{align}
with "con" indicating the connected correlation function. We show below that
\begin{align}
 \label{Ex1} \langle \phi_i\rangle_0 &=-\sum_a \sum_j J_a M_{aj}(M^{-1})_{ji},\\
 \label{Ex2} \langle \phi_i\phi_j\rangle_0 &=(M^{-1})_{ij}+\langle\phi_i\rangle_0\langle \phi_j\rangle_0,\\
 \label{Ex3} \langle\phi_i\phi_j\rangle_{0,\rm con} &= (M^{-1})_{ij},
\end{align}
whereas $\langle \phi_i^3\rangle_{0,\rm con}=0$. We then obtain
\begin{align}
 w_a^{(i)} &= -3 M_{ii} \sum_j M_{aj}G_{ji},\\
 w_{abc}^{(i)} &=- \Bigl(\sum_{j} M_{aj} G_{ji}\Bigr)\Bigl(\sum_{k} M_{bk} G_{ki}\Bigr)\Bigl(\sum_{l} M_{cl} G_{li}\Bigr),
\end{align}
the latter being symmetric in the indices $a,b,c$. The perturbative generating function becomes
\begin{align}
 \nonumber Z \simeq{}& e^{-\frac{u}{3!}\sum_a J_a^3}Z_0\Bigl(1-\frac{u}{3!} \sum_i\sum_a J_a w_a^{(i)} \\
 &-\frac{u}{3!}\sum_i\sum_{a,b,c} J_aJ_bJ_c w_{abc}^{(i)}\Bigr).
\end{align}
We re-exponentiate this to obtain
\begin{align}
 \nonumber \ln Z \simeq{}& -\frac{1}{3!}\sum_a u J_a^3 - S_{0,\rm eff} -\frac{u}{3!} \sum_i\sum_a J_a w_a^{(i)} \\
 &-\frac{u}{3!}\sum_i\sum_{a,b,c} J_aJ_bJ_c w_{abc}^{(i)}.
\end{align}
We then find
\begin{align}
 \langle \mathcal{O}_a\mathcal{O}_b\mathcal{O}_c\rangle &= \frac{\partial^3\ln Z}{\partial J_a\partial J_b\partial J_c}  =-u \delta_{ab}\delta_{ac}- u \sum_i w_{abc}^{(i)}.
\end{align}
For $a\neq b$ we arrive at
\begin{align}
\nonumber &\langle \mathcal{O}_a\mathcal{O}_b\mathcal{O}_c\rangle \\
&= u\sum_i\Bigl(\sum_{j} M_{aj} G_{ji}\Bigr)\Bigl(\sum_{k} M_{bk} G_{ki}\Bigr)\Bigl(\sum_{l} M_{cl} G_{li}\Bigr),
\end{align}
which is Eq. (8).

\emph{Correlation functions of $\phi$ with sources.} We now derive Eqs. (\ref{Ex1})-(\ref{Ex3}). To compute correlation functions of bulk fields $\phi_i$ in the presence of boundary sources $J_a$, we introduce auxiliary bulk sources $h_i$ according to
\begin{align}
 &Z_0(\{J_a,h_i\}) = \int\Bigl(\prod_i \mbox{d}\phi_i\Bigr) e^{-S_0(\{\phi_i,J_a\})+\sum_ih_i\phi_i}\\
 &= e^{-S_{0,\rm eff}[J]-\sum_{i,j}\sum_aM_{ai} J_a(M^{-1})_{ij}h_j+\frac{1}{2}\sum_{i,j}(M^{-1})_{ij}h_ih_j}
\end{align}
and correlation functions are obtained from
\begin{align}
 \langle \phi_{i_1}\cdots\phi_{i_n}\rangle_0 &= \frac{\partial^n Z_0}{\partial h_{i_1}\cdots \partial h_{i_n}}\Bigr|_{h=0},\\
  \langle \phi_{i_1}\cdots\phi_{i_n}\rangle_{0,\rm con} &= \frac{\partial^n\ln Z_0}{\partial h_{i_1}\cdots \partial h_{i_n}}\Bigr|_{h=0}.
\end{align}
This yields Eqs. (\ref{Ex1})-(\ref{Ex2}). All connected $n$-point functions of $\phi_i$ vanish, because $\ln Z_0$ is quadratic in $h$.

\section{S6. Derivation of Eq. (10)} \label{SecGeo}

In this section we discuss the expected behavior of the 2-pt function $\langle \mathcal{O}_a\mathcal{O}_b\rangle$ for both type-I and type-II geometries in the continuum. This enables the identification of the respective expressions for $d_{ab}$ in Eq. (10). Our approach is based on the observation that the two-point function 
asymptotically behaves like $\langle \mathcal{O}_a\mathcal{O}_b\rangle \simeq e^{-\Delta \sigma_{ab}/\ell}$, where $\sigma_{ab}$ is the large geodesic distance between two sites $z_a$ and $z_b$. The latter, however, can be computed  in the continuum for any two sites $z$ and $z'$.

Both type-I and type-II geometries are locally AdS, with a Riemannian (all-positive) signature. The equation of motion of real scalar fields in the bulk are expressed through the Laplace--Beltrami operator   $\square=(1/\sqrt{\mbox{det}(g)})\partial_\mu (\sqrt{\mbox{det}(g)}g^{\mu\nu}\partial_\nu)$. The bulk Green function satisfies
  \beq
  \left(-\square+m^2\right)G(Z;Z')=\frac{1}{\sqrt{\mbox{det}(g)}}\delta^{(d+1)}(Z-Z').
  \eeq
where $Z$ denotes the generalized coordinate of the $\rm AdS_{2}$ with   $d=1$. For general $\rm AdS_{d+1}$, the solution is
\begin{align}
   \label{AdSGreens}G(Z;Z')=&G(\xi(Z,Z')),\\
 \nonumber   G(\xi) =&
 \xi^{\Delta}\ {}_2F_1\Bigl(\Delta,\frac{d}{2};
  \Delta+1-\frac{d}{2};\xi\Bigr),
\end{align}
 where $m^2\ell^2=\Delta(\Delta-d) $, and $0\leq \xi(Z,Z')^{-1}=\cosh(\sigma(Z,Z')/\ell)\leq 1$ is a dimensionless 
 parameter, determined by the geodesic distance $\sigma(Z,Z')$ \cite{ammon15}. 
 The hypergeometric function is denoted by ${}_2F_1$. For large $\sigma$, 
 the term containing ${}_2F_1$ can be neglected and we have $G(\sigma)\simeq \xi^\Delta \simeq e^{-\Delta \sigma/\ell}$.

\emph{Type-I geometry.}---The geodesic distance in the Poincar\'{e} disk with metric
\begin{align}
\mbox{d}s^2_{\rm I} = \frac{(2\ell)^2}{(1-|z|^2)^2}|\mbox{d}z|^2
\end{align}
is given by
\begin{align}
 \sigma(z,z') = \ell\ \text{arcosh}\Bigl(1+\frac{2|z-z'|^2}{(1-r^2)(1-r'{}^2)}\Bigr).
\end{align}
For sites close to the boundary, $r,r'\to 1$, we can use $\text{arcosh}(x)\simeq \ln(2x)$ for $x\gtrsim 4$, to approximate this as
\begin{align}
 \sigma(z,z') \simeq \ell\ \ln\Bigl(\frac{4|z-z'|^2}{(1-r^2)(1-r'{}^2)}\Bigr).
\end{align}
The corresponding correlation function reads
\begin{align}
 G(z;z') \simeq \Bigl(\frac{(1-r^2)(1-r'{}^2)}{4|z-z'|^2}\Bigr)^{\Delta}.
\end{align}
For two sites $z=r e^{\rmi \theta}$ and $z'=r'e^{\rmi \theta'}$ close to the boundary, we then obtain the boundary 2-pt function
\begin{align}
\label{Gone} G(\theta,\theta') \simeq \frac{\tilde{C}_2}{|e^{\rmi \theta}-e^{\rmi \theta'}|^{2\Delta}}.
\end{align}
Comparing this with
\begin{align}
 G(\theta_a,\theta_b) \simeq \frac{\bar{C}_2}{(d_{ab})^{2\Delta}}
\end{align}
leads to the first line of Eq. (10).

The coefficient $\tilde{C}_2$ depends on $r,r'$ and, technically, vanishes as $r,r'\to 1$. In the continuum, choosing the boundary on a circle of radius $R<1$, this can be absorbed by a redefinition of the field, while on the discrete lattice (due to the variation in the discrete $r_i$), there is a residual dependence of $\langle\mathcal{O}_a\mathcal{O}_b\rangle$ on $r_a$ and $r_b$. This effect, however, is small for large enough flakes. 

\emph{Type-II geometry.} The continuum geometry of the type-II lattice (Eq. 4 and Eq. \ref{EqS18}) can be parametrized
in angular coordinates $\hat z = \hat r e^{i\theta}$ through the metric
\beq
\mbox{d}s^2_{\rm II} &= \left(\frac{\ell w}{4}\right)^2\frac{\mbox{d}\hat{r}^2+\hat{r}^2\mbox{d}\theta^2}{\hat{r}^2\sin^2\left(\frac{w}{4}\log\hat{r}\right)}\\
&= \ell^2\frac{\mbox{d}[\frac{w}{4}\log\hat{r}]^2+\mbox{d}[\frac{w}{4}\theta]^2}{\sin^2\left(\frac{w}{4}\log\hat{r}\right)},
\label{typeII-angMetric}
\eeq 
with $w=kP=8\pi T\ell$ as defined in Eq. (\ref{EqS45}). The locally-AdS geometry can also be described by the upper half-plane 
metric
\beq
\mbox{d}s^2_{\rm II} =\frac{\ell^2}{y^2}\left(\mbox{d}x^2+\mbox{d}y^2\right),
\eeq 
from which the angle-coordinate metric [Eq.~\eqref{typeII-angMetric}] of the same geometry is obtained
by the coordinate transformations,
\beq
x&=e^{w\theta/4}\cos\Bigl(\frac{w}{4}\log\hat r\Bigr),\\
y&=e^{w\theta/4}\sin\Bigl(\frac{w}{4}\log\hat r\Bigr).
\label{type-II-to-uhp}
\eeq 
The coordinate transformation also provides a map to 
the type-II Green function from the corresponding function expressed 
in terms of the upper half-plane geodesic
distances. For two sites  on the Poincar\'{e} upper half-plane, $(x,y)$ and $(x',y')$, the Green function for long distances
is given by \cite{ammon15}
\beq
G(x,y;x',y')\simeq \left(\frac{2yy'}{y^2+{y'}^2+(x-x')^2}\right)^\Delta.
\eeq 
Define $\chi=\frac{w}{4}\log\hat r$, hence $(x,y)= e^{w\theta/4}(\cos\chi,\sin\chi)$,  so that  the Green function becomes
\begin{widetext}
\beq
&G(\chi,\theta;\chi',\theta') \simeq \Biggl(\frac{2\sin\chi\sin\chi'e^{\frac{w}{4}(\theta+\theta')}}{e^{\frac{w}{2}\theta}\sin^2\chi+{e}^{\frac{w'}{2}\theta'}\sin^2\chi'+(e^{\frac{w}{4}\theta}\cos\chi-e^{\frac{w}{4}\theta'}\cos\chi')^2}\Biggr)^\Delta.
\eeq 
\end{widetext}
To approach the outer boundary of the type-II graph, $\hat{r}\to 1$, we introduce an infrared regulator $\varepsilon$ according to $\hat{r} = e^{-4\pi \vare/w}$ or, equivalently, $\chi =-\pi \vare$. The two-point function on the boundary becomes
\beq
&G(\theta,\theta')
\simeq &\lim_{\varepsilon,\varepsilon'\rightarrow 0}
\Biggl(\frac{2\pi^2\varepsilon\varepsilon'}{[\sinh{\pi T\ell (\theta-\theta')}]^2}\Biggr)^\Delta.
\eeq 
As in the type-I case, the vanishing prefactor as $\vare,\vare'\to 0$ can be absorbed into a redefinition of the boundary field.
By enforcing angular periodicity via the method of images, we obtain the 2-point function between
between boundary sites $\theta_a$ and $\theta_b$ as
\beq
G(\theta_a,\theta_b)\simeq \bar{C}_2 \sum_{n\in\mathbb{Z}} \Bigl(\frac{1}{\sinh[\pi T \ell(\theta_a-\theta_b+2\pi n)]}\Bigr)^{2\Delta}.
\eeq 
This form validates the choice of $d_{ab}$ for the type-II graphs in Eq. (10).

\section{S5. Derivation of Eq. (11)}\label{SecMass}

To derive Eq. (11), we extend the presentations in Refs. \cite{PhysRevA.102.032208,Stegmaier2021}. We consider the non-interacting part of the lattice action $S(\{\phi_\mu\})$ and consider the equation of motion for the field on a bulk site $z_i\in\mathbb{D}$ with $q$ neighbors given by
\begin{align}
   -t\sum_\mu A_{\mu i} \phi(z_i) + \hat{m}^2 \phi(z_i) =0.
\end{align}
We set $t=1$. This can be written as
\begin{align}
 \label{h1a} \sum_{\alpha=1}^q \phi\Bigl(\frac{z_i-w_\alpha}{1-w_\alpha \bar{z}_i}\Bigr) = \hat{m}^2 \phi(z_i),
\end{align}
where the sum extends over the neighbors of $z_i$. Here $w_\alpha= h e^{2\pi(\alpha-1)\rmi/q}e^{\rmi \chi_i}$ and $\chi_i$ a defect angle that depends on $z_i$. The dimensionless parameter $h$ is a lattice constant given by
\begin{align}
 h = \Bigl(1-\frac{\sin^2(\frac{\pi}{q})}{\cos^2(\frac{\pi}{p})}\Bigr)^{1/2}.
\end{align}
The expression on the left-hand side of Eq. (\ref{h1a}) can be expanded in powers of $h$ with the universal leading term given by \cite{PhysRevA.102.032208}
\begin{align}
 \label{h1} \Bigl(q + q h^2 \ell^2 \square+O(h^3)\Bigr)\phi(z_i) = \hat{m}^2 \phi(z_i).
\end{align}
This is equivalent to the Klein--Gordon type equation
\begin{align}
 (-\square+m^2)\phi(z) =0
\end{align}
through the identification
\begin{align}
 \hat{m}^2 = q + qh^2m^2\ell^2.
\end{align}

We define the dimensionless Laplacian $\hat{\square}=\ell^2\square$. In the Poincar\'{e} disk with $z=x+\rmi y \in\mathbb{D}$ we have $\hat{\square}=\frac{1}{4}(1-|z|^2)^2(\partial_x^2+\partial_y^2)=(1-|z|^2)^2\partial_z\bar{\partial}_z$. Eigenfunctions of $\hat{\square}$ are labelled by a real quantum number $\Delta$ such that $\hat{\square} \phi_\Delta(z) = \Delta(\Delta-1)\phi_\Delta(z)$. (This is most easily seen in upper half-plane coordinates $x+\rmi y \in\mathbb{H}$ with $\hat{\square} = y^2(\partial_x^2+\partial_y^2)$ and $\phi_\Delta(x,y)=y^\Delta$.) We have
\begin{align}
\Bigl(q + qh^2 \frac{\Delta(\Delta-1)}{\ell^2}+O(h^3)\Bigr)\phi_\Delta(z_i) = \hat{m}^2 \phi_\Delta(z_i),
\end{align}
or, 
\begin{align}
\Delta(\Delta-1)+ O(h) = m^2\ell^2,
\end{align}
which recovers the universal continuum limit result.

Higher orders in an expansion in the parameter $h$ in Eq. (\ref{h1}) can be derived. Importantly, terms that are not powers of $\hat{\square}$ appear at order $h^q$, and so the power series needs to be stopped earlier. For the $\{3,7\}$ tessellation with $q=7$ we have
\begin{align}
&q\Biggl(1+h^2\hat{\square}+\frac{h^4}{4}(\hat{\square}^2+2\hat{\square})\\
&+\frac{h^6}{36}(\hat{\square}^3+10\hat{\square}^2+12\hat{\square})+O(h^7)\Biggr)\phi(z_i) = \hat{m}^2\phi(z_i),
\end{align}
which is of the form
\begin{align}
 &q + qh^2 f(\hat{\square})\phi(z_i) = \hat{m}^2\phi(z_i)
\end{align}
with $f(X)$ given in Eq. (10). Acting on eigenfunctions $\phi_\Delta$ we obtain
\begin{align}
 f(\Delta(\Delta-1))\phi_\Delta = m^2\ell^2\phi_\Delta,
\end{align}
and hence the lattice relation between $\Delta$ and  $m^2\ell^2$. For a general $\{p,q\}$ lattice, the expansion reads
\begin{align}
 f(X) = \sum_{l=1}^{l_{\rm max}} \frac{h^{2(l-1)}}{l!^2}T_l(X) + O(h^{q-2}),
\end{align}
which has to be stopped at order $l_{\rm max}$ such that $2(l_{\rm max}-1)< q-1$. For instance $q=3$ means $l_{\rm max}=1$ and $q=7$ means $l_{\rm max}=3$. The polynomials $T_l(X)$ are
\begin{align}
 T_1(X) &= X,\\
 T_2(X) &= X^2+2X,\\
 T_3(X) &= X^3 + 10X^2+12X,\\
 T_4(X) &= X^4 + 28X^3 + 156X^2+144X.
\end{align}
They can be computed systematically to arbitrary order from a Taylor expansion of Eq. (\ref{h1a}).

\section{S7. Fit procedure}\label{SecFit}

\subsection{Type-I / Zero temperature}\label{zeroTfit}

\begin{figure}[tb!]
  \includegraphics[width=\linewidth]{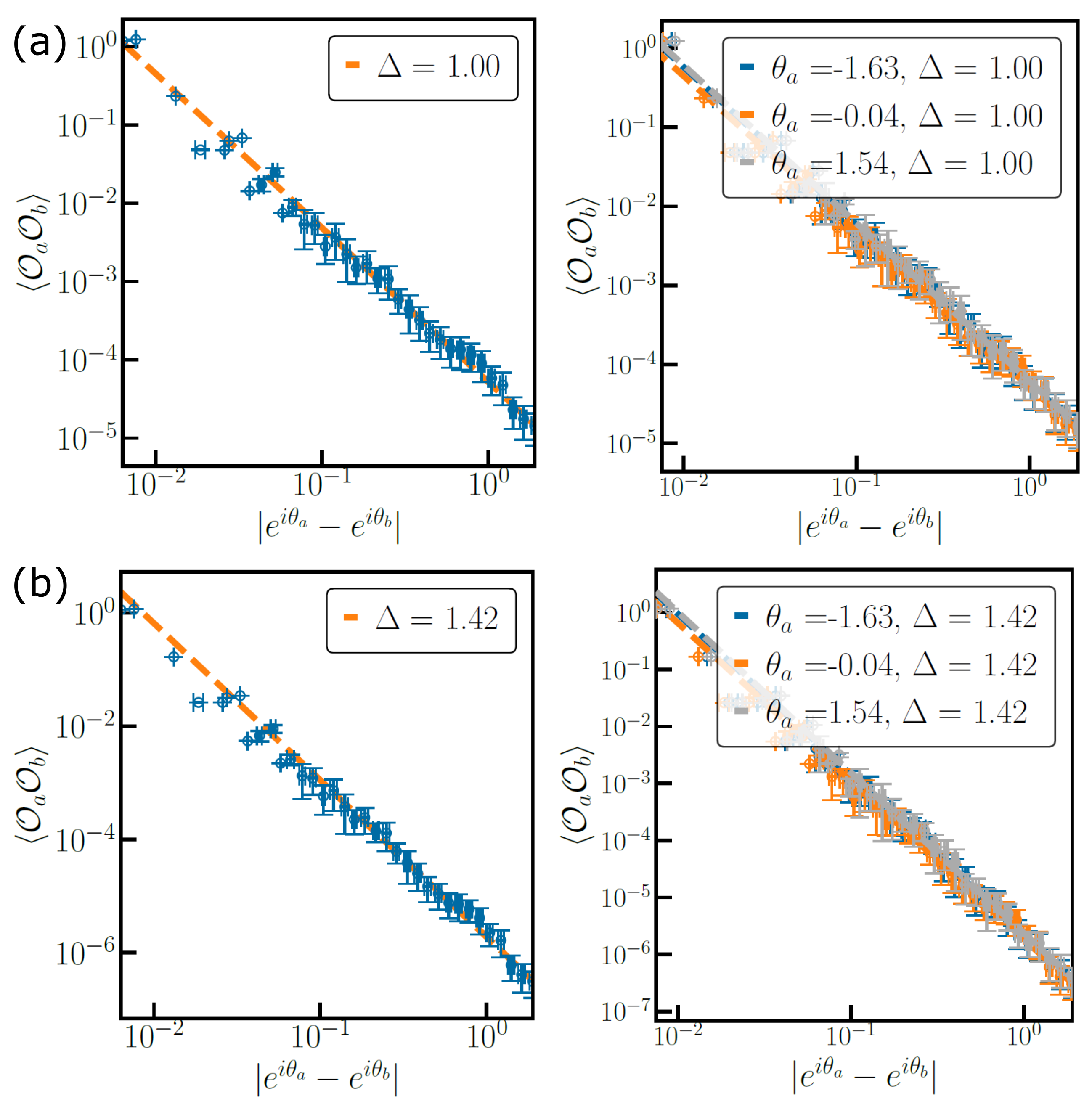}
  \caption{The power-law scaling of the 2-point boundary correlation functions obtained from Eq. (7), between two boundary
  points $a$ and $b$. In \textbf{(a)} and \textbf{(b)}, the function on a type-I flake with 6 shells are shown 
  with bulk scalar masses $m^2\ell^2=0.01$ and $0.74$, respectively. The quoted values of $\Delta$ are the fitted values. On the left panels, 
  a fixed boundary site at $\theta_a=0$ is chosen as a reference.
  On the right panels, three separate 
  angular positions have been chosen for $a$, as the site $b$ crawls along the boundary. The discreteness of the
  boundary introduces small amplitude variations but the scaling dimension remains largely 
  unaffected, demonstrating the expected CFT behavior at zero temperature. 
  The scatter (depicted as error bars) results from the binning of the boundary distances. 
  Note that $m=0$ is consistent with $\Delta=1$.
  }
  \label{2pt_analysis}
\end{figure}

\emph{2-point functions.}---We first check the scaling form
of the 2-pt functions on the boundary by tuning the bulk scalar mass. In Fig.~\ref{2pt_analysis},
we verify the scaling behavior
\begin{align}
  \langle \mathcal{O}_a\mathcal{O}_b\rangle \simeq \frac{\bar{C}_2}{|e^{\rmi \theta_a}-e^{\rmi \theta_b}|^{2\Delta}}
\end{align}
for $\langle\mathcal{O}_a\mathcal{O}_b\rangle$ obtained from Eq. (7) in the main text, with the 
two boundary points $\theta_a$ and $\theta_b$ on the Poincar\'{e} 
disk. Fig.~\ref{2pt_analysis} uses a binning of the
correlation function
dataset to reduce small-distance discretization effects. 
We consider different sample sizes
for the binning 
and perform a least-square fitting of the data on a logarithmic 
grid. We pick the binning with the least net residual for each dataset depicted in Fig.~\ref{2pt_analysis}~(a). 
While the binning reduces the scatter in the local magnitudes of the correlation function, the 
discrete lattice
also introduces a ruggedness in the boundary geometry that requires further consideration.
However, this effect is mitigated by sampling over several distinct pairs of boundary locations 
that share the same global distances. In effect, we fix the location of one site $a$, and let the 
other site $b$ move all the way across the compact boundary. The right-hand panels in Fig.~\ref{2pt_analysis}
depict the variation of the 2-point function for three such choices of
the anchored site $a$. By averaging through many such boundary reference frames 
distinguished by the anchoring of their origin coordinates $a$, we obtain
the mass parametrization of the scaling dimension $\Delta$ that is depicted in the main text 
along with its related standard deviation.

\begin{figure}[tb!]
  \includegraphics[width=\linewidth]{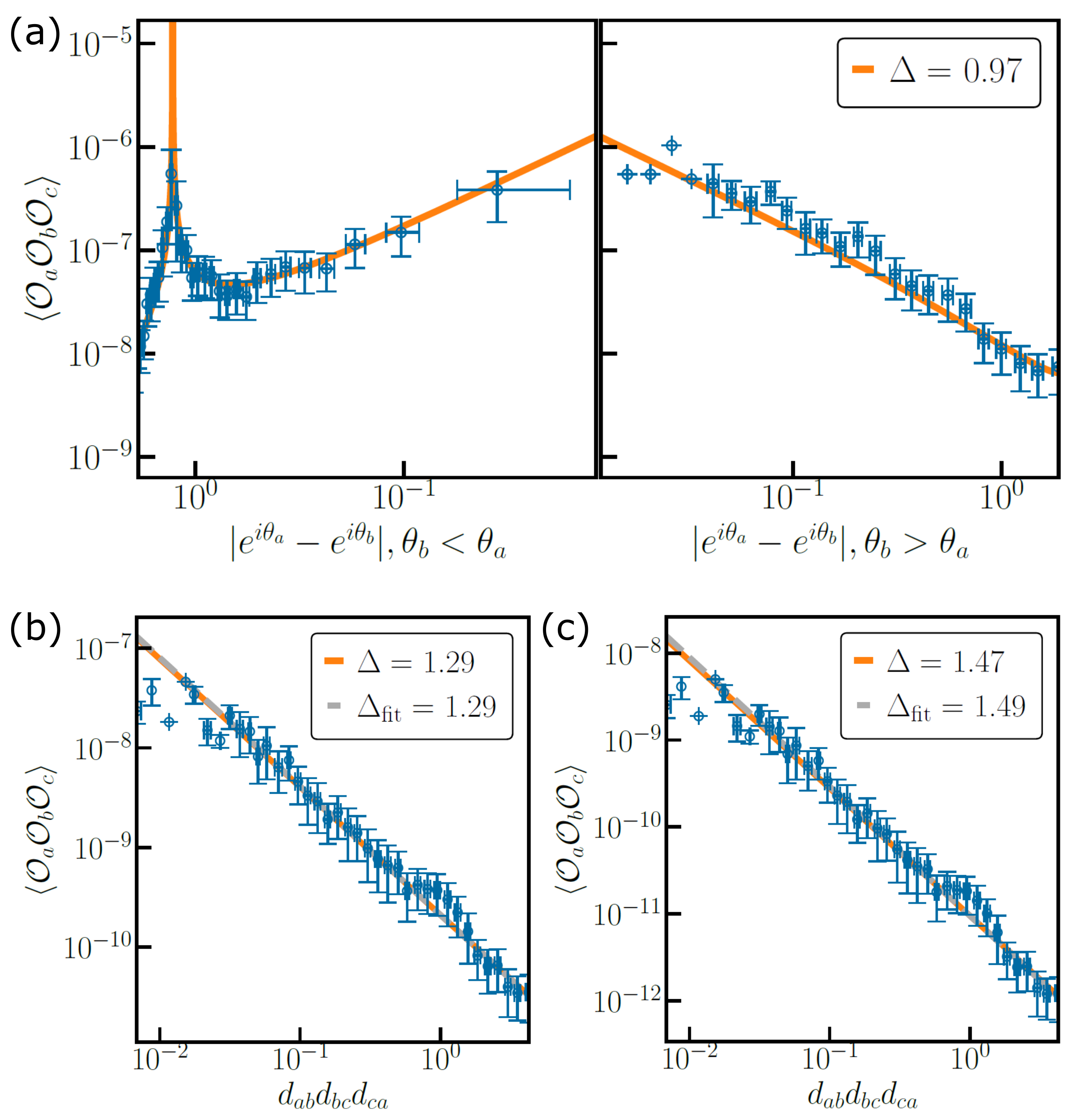}
  \caption{The conformal scaling of the 3-point function computed from
  Eq. (8), applied to type-I graphs with 7 shells for masses $m^2\ell^2=0.01$, $0.48$, and $0.90$, plotted 
  respectively in \textbf{(a)}, \textbf{(b)}, and \textbf{(c)}. The figures depict the correlation function for 
  two fixed boundary sites $a$ and $c$, with a third site 
  $b$ that moves along the boundary. In \textbf{(a)}, the function is plotted against the relative distance between
  the $a$ and $b$ coordinates, depicting the contact divergence
  when $b$ coincides with the third coordinate $c$. 
  In \textbf{(b)} and \textbf{(c)}, a scaling collapse over the product of the pairwise relative distances, 
  fitted as the dashed grey lines,
    shows a remarkable agreement with the solid orange lines defined by the scaling dimension obtained from the 2-point
    function.
    }
  \label{3pt_analysis}
\end{figure}

\emph{3-point functions.}---For the 3-point function, we use Eq. (8) to
obtain the boundary correlators and compare to the zero temperature scaling formula
\begin{align}
 \langle \mathcal{O}_a\mathcal{O}_b\mathcal{O}_c\rangle \simeq \frac{\bar{C}_3}{(d_{ab}d_{ac}d_{bc})^\Delta}
\end{align}
with $d_{ab}=|e^{\rmi \theta_a}-e^{\rmi \theta_b}|$. Similar to the 2-point function, a linear least-square
fit of the 3-point function over the product over the three relative distances $d_{ab}d_{ac}d_{bc}$ 
demonstrates the expected power-law behavior as seen in Fig.~\ref{3pt_analysis}. In Fig.~\ref{3pt_analysis}~(b)
and (c), a robust
evidence for the conformal symmetry emerges as the scaling dimension $\Delta$ obtained from the scaling-collapse of the 3-point function
for different masses shows good agreement with $\Delta$ obtained from the 2-point function. 
To average out the discreteness of the data, we consider independent sets of 
3-point functions with one fixed angle $\theta_a$,
and two varying angles $\theta_b$ and $\theta_c$, the last of which parametrizes the data sets. A given realization
with a fixed set of two angles $\theta_{a,c}$ (Fig.~\ref{3pt_analysis} shows an example) contains 3-point function data
that is binned and fitted, and the overall scaling dimension is obtained from averaging over the different data sets. This yields the comparison between the 2-point and 3-point fits of the scaling dimension shown in Fig 3d in the main text.

\subsection{Type-II / Finite temperature}

\begin{figure}[tb!]
  \includegraphics[width=\linewidth]{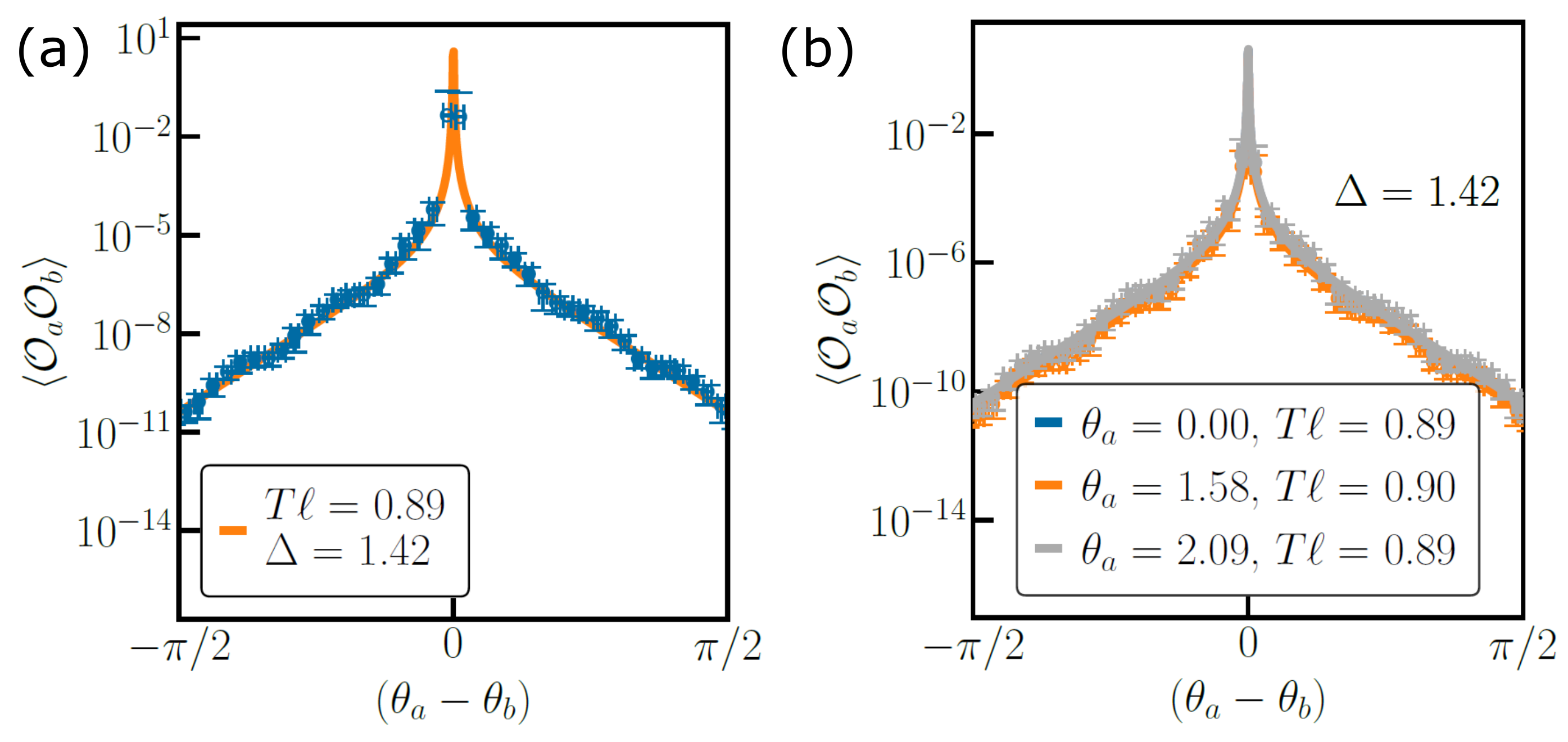}
  \caption{
    Thermal boundary 2-point function obtained by applying Eq. (7) for a type-II graph with $k=12$, 
    obtained by squash-and-wrap from a flake with 6 shells, and 
    bulk scalar mass $m^2\ell^2 = 0.740$. \textbf{(a)} The orange fit line is given by Eq.~\ref{therm2pt} 
    with the sum extending over $n\in[-120,120]$. In (\textbf{d}), the variation of the 2-point function over a 
  set of varying anchoring coordinates $\theta_a$ is shown to lead to only a mild variation of the fitted temperature.}
  \label{T_compare}
\end{figure}

\begin{figure*}[tb!]
  \includegraphics[width=\linewidth]{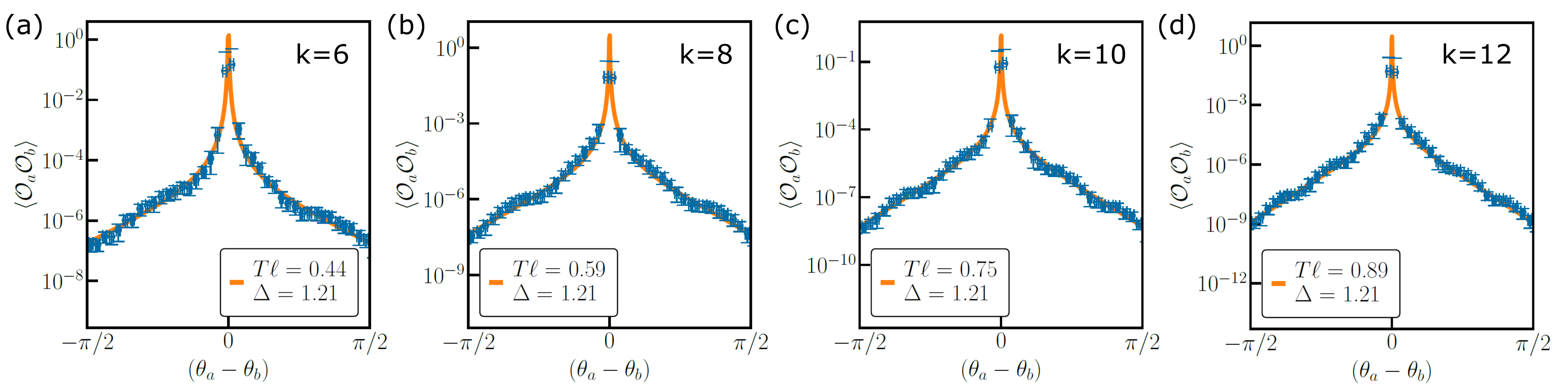}
  \caption{Type-II 2-point functions for $m^2\ell^2=0.32$ shown for   (\textbf{a}) $k=6$,  (\textbf{b}) $k=8$,  (\textbf{c}) $k=10$,
  and,  (\textbf{d}) $k=12$,  plotted vs. $\theta_a-\theta_b$ along with orange fit lines from Eq.~\eqref{therm2pt}.
  }
  \label{2pt_analysis_finiteT}
\end{figure*}

\begin{figure*}[tb!]
  \includegraphics[width=\linewidth]{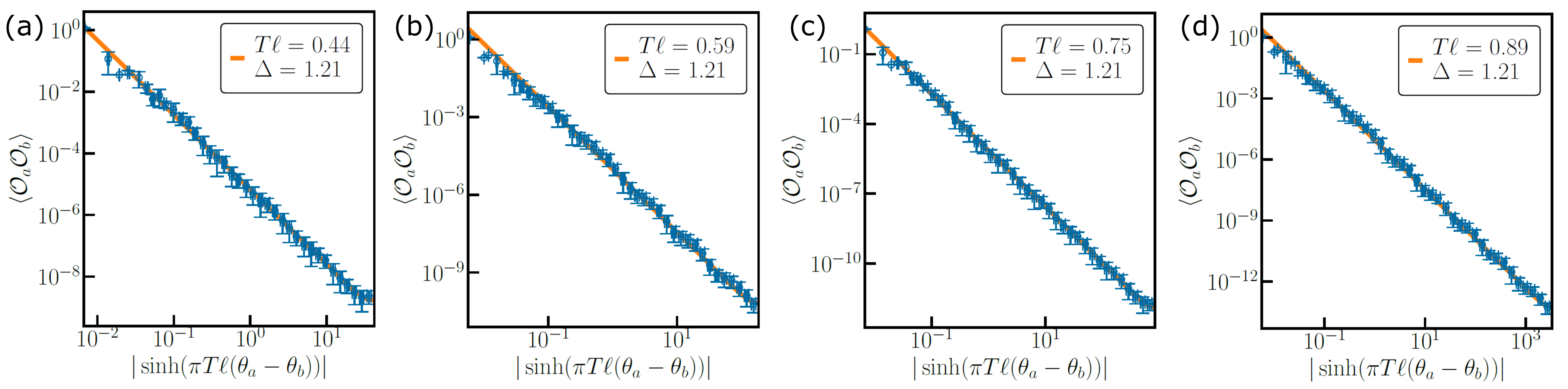}
  \caption{Type-II 2-point functions for $k=6,8,10$, and $12$ (\textbf{(a)-(d)}) as in Fig. \ref{2pt_analysis_finiteT}, plotted vs. $d_{ab}(T)$.}
  \label{2pt_analysis_finiteT_loglog}
\end{figure*}

\emph{2-point functions.}---We compute the 2-pt function through Eq. (7) on the type-II flakes for various values of $k$. To extract the temperature of the boundary CFT, we fit the boundary 2-pt correlator to the
continuum limit form 
\begin{align}
\label{therm2pt} \langle \mathcal{O}_a\mathcal{O}_b\rangle = \sum_{n\in\mathbb{Z}} \frac{\bar{C}_2'}{(d_{ab}^{(n)})^{2\Delta}}
\end{align}
with 
\begin{align}
 d_{ab}^{(n)}= \frac{\sinh(\pi T \ell |\theta_a-\theta_b+2\pi n|)}{\pi T \ell}.
\end{align}
This continuum form is derived from the formula for a constant time-slice of the 3D BTZ black hole geometry. The sum over $n$ is reminiscent of the method of images, and produces a periodic function. 
We find that the sum converges quickly with $|n|\lesssim 120$ images and yields a very good agreement, see Fig.~\ref{T_compare}.
While the scaling dimension $\Delta(m^2\ell^2)$ can be fitted for the type-II 2-pt function, we find that it is consistent with the type-I result $\Delta(m^2\ell^2)$. For this reason, to increase the accuracy of the temperature fit, we use the relation $\Delta(m^2\ell^2)$ from type-I for the fit of $T\ell$ in Eq. (\ref{therm2pt}) and confirm the relation $T\ell = kP/8\pi$.
The rugged-boundary problem is addressed in the same way as for the type-I graphs discussed above. We average the
temperature fits 
across various $\theta_a$-parametrized data sets, leading to the favorable comparison with the Hawking temperature
associated with the BTZ geometry.

\begin{figure}[tb!]
  \includegraphics[width=\linewidth]{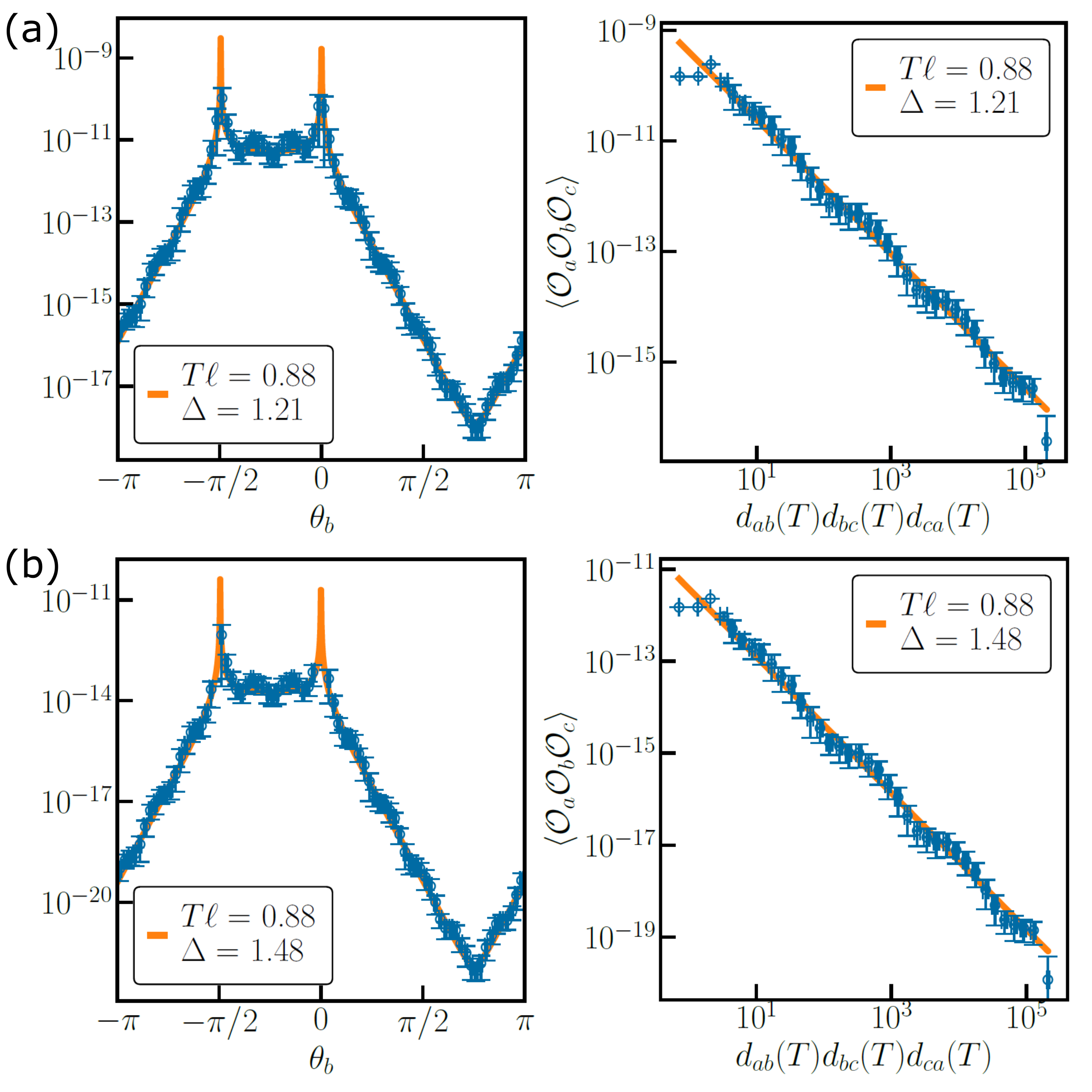}
  \caption{The 3-point functions at the boundary of the type-II graph with $k=12$,
  for $m^2\ell^2=0.32$ \textbf{(a)} and $m^2\ell^2=0.90$ \textbf{(b)}.
  The observed contact divergences (left panels) and asymptotic
  thermal scaling (right panels) are consistent with the expected conformal behavior.
  }
  \label{3pt_analysis_finiteT}
\end{figure}

\emph{3-point functions.}---To produce Fig 3b in the main text, we fix $\theta_a=0$ and $\theta_c$, and vary $\theta_b$. We compare the 3-point function obtained from applying Eq. (8) against the form anticipated for a timeslice of a 3D BTZ black hole given by
  \begin{align}
  \label{therm3pt} \langle \mathcal{O}_a\mathcal{O}_b\mathcal{O}_c\rangle &=
  \sum_{n\in\mathbb{Z}} \left.\frac{\bar{C}_3'}{(d_{ab}d_{ac}d_{bc})^{\Delta}}\right|_{\theta_b\rightarrow\theta_b+2\pi n},\\
  &=
  \sum_{n\in\mathbb{Z}} \frac{\bar{C}_3'}{(d_{ab}^{(n)}d_{ac}d_{bc}^{(-n)})^{\Delta}}.
  \end{align}
The case of various masses is shown in Fig. \ref{3pt_analysis_finiteT}, together with 
with their scaling collapse.

\begin{figure}[tb!]
  \includegraphics[width=\linewidth]{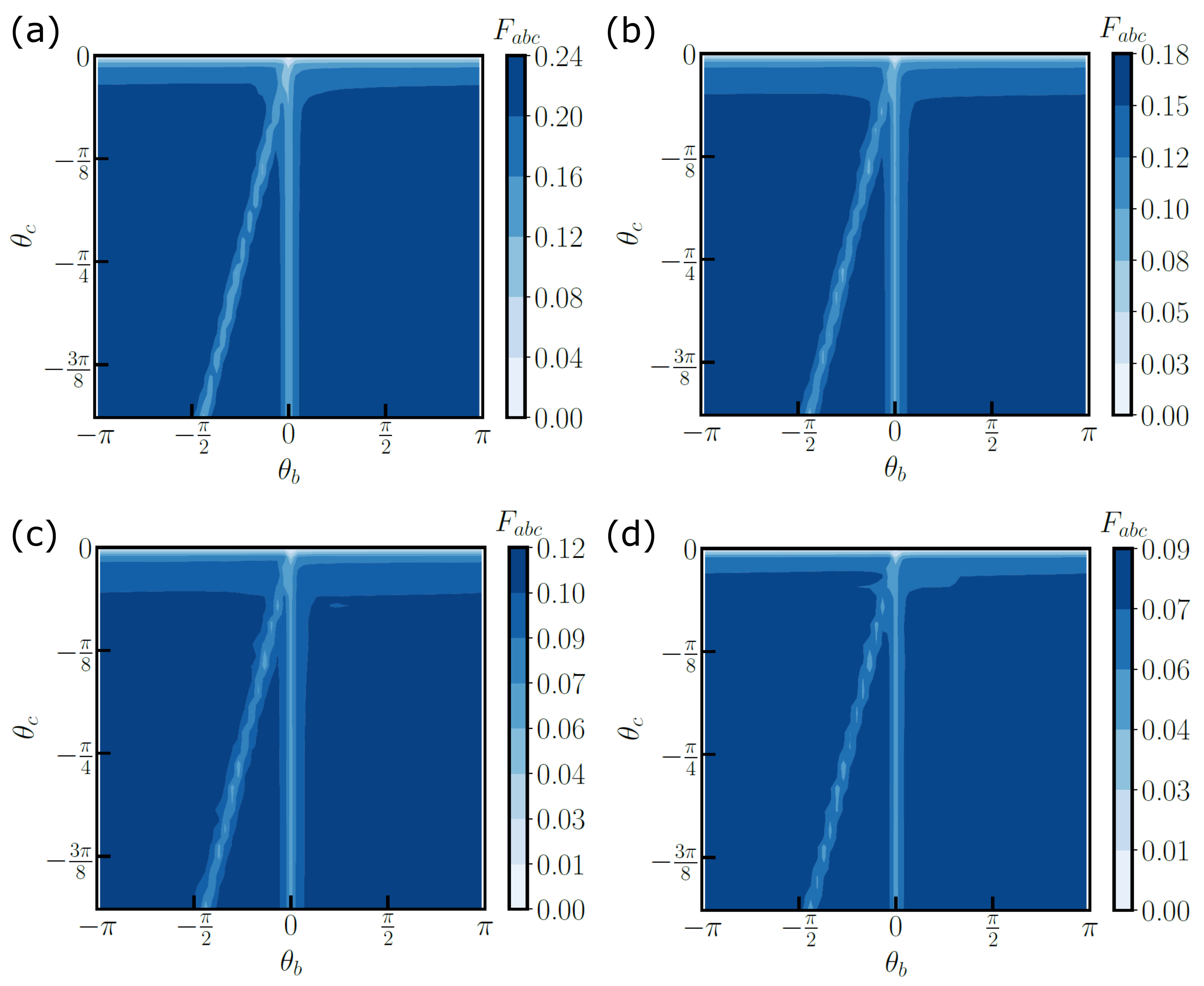}
  \caption{
    Normalized 3-point ratio (\ref{FabcEq}) plotted for fixed $\theta_a = 0$, and 
    varying values of $\theta_b$ and $\theta_c$ on the boundary of the type-I graph 
    obtained for the bulk scalar mass $m^2\ell^2=0.11$ \textbf{(a)}, $0.32$ \textbf{(b)}, $0.63$ \textbf{(c)}, 
    and $0.95$ \textbf{(d)}. The value of $F_{abc}$ approaches the saturation value $C_3$ away
    from the lines the contact divergences, given by $\theta_c=\theta_a=0$,
    $\theta_b=\theta_a=0$, and $\theta_c=\theta_c$.}
  \label{FabcFig}
\end{figure}

\subsection{3-pt coefficient}\label{BCFT}
The normalized 3-point coefficient $C_3$ is given by
\begin{align}
 C_3 = \frac{\bar{C}_3}{(\bar{C}_2)^{3/2}}
\end{align}
in the type-I case, and similarly $C_3 = \bar{C}_3'/(\bar{C}_2')^{3/2}$ in the type-II case. For a systematic way to fit $C_3$, we construct the ratio
\begin{align}
  \label{FabcEq}
 F_{abc} = \frac{\langle \mathcal{O}_a\mathcal{O}_b\mathcal{O}_c\rangle}{[\langle\mathcal{O}_a\mathcal{O}_b\rangle\langle\mathcal{O}_a\mathcal{O}_c\rangle\langle\mathcal{O}_b\mathcal{O}_c\rangle]^{1/2}}
\end{align}
that we sample over many values of $a,b,c$. 
 The idea of using $F_{abc}$ is that, naively, one could simply sample
\begin{align}
   \lim_{d_{ab}d_{bc}d_{ca}\to\infty}\langle \mathcal{O}_a\mathcal{O}_b\mathcal{O}_c\rangle(d_{ab}d_{ac}d_{bc})^\Delta \simeq C_3,
\end{align}
and determine $C_3$ from this. However, for the type-II lattices we construct here as described in Sec. S2, the ring boundary is too rough and leads to a $k$-fold modulation of the 2-pt function already, which amplifies when considering the 3-pt function.
(The type-I boundaries are also somewhat rough but less so than the type-II boundaries.)
This could be incorporated by appropriate form-factors for both the 2-pt and 3-pt function. By considering $F_{abc}$, instead, these effects are expected to cancel, and eventually yield a cleaner prediction of the actual physical type system, whose discrete boundary can be made as smooth as possible, in principle.
As Fig.~\ref{FabcFig} shows, $F_{abc}$ only mildly depends on $a,b,c$ and can be 
used to determine $F_{abc} \simeq C_3$  as long as we stay away from the contact divergences that arise when any two of the boundary sites
coincide.

\begin{figure}[tb!]
  \includegraphics[width=\linewidth]{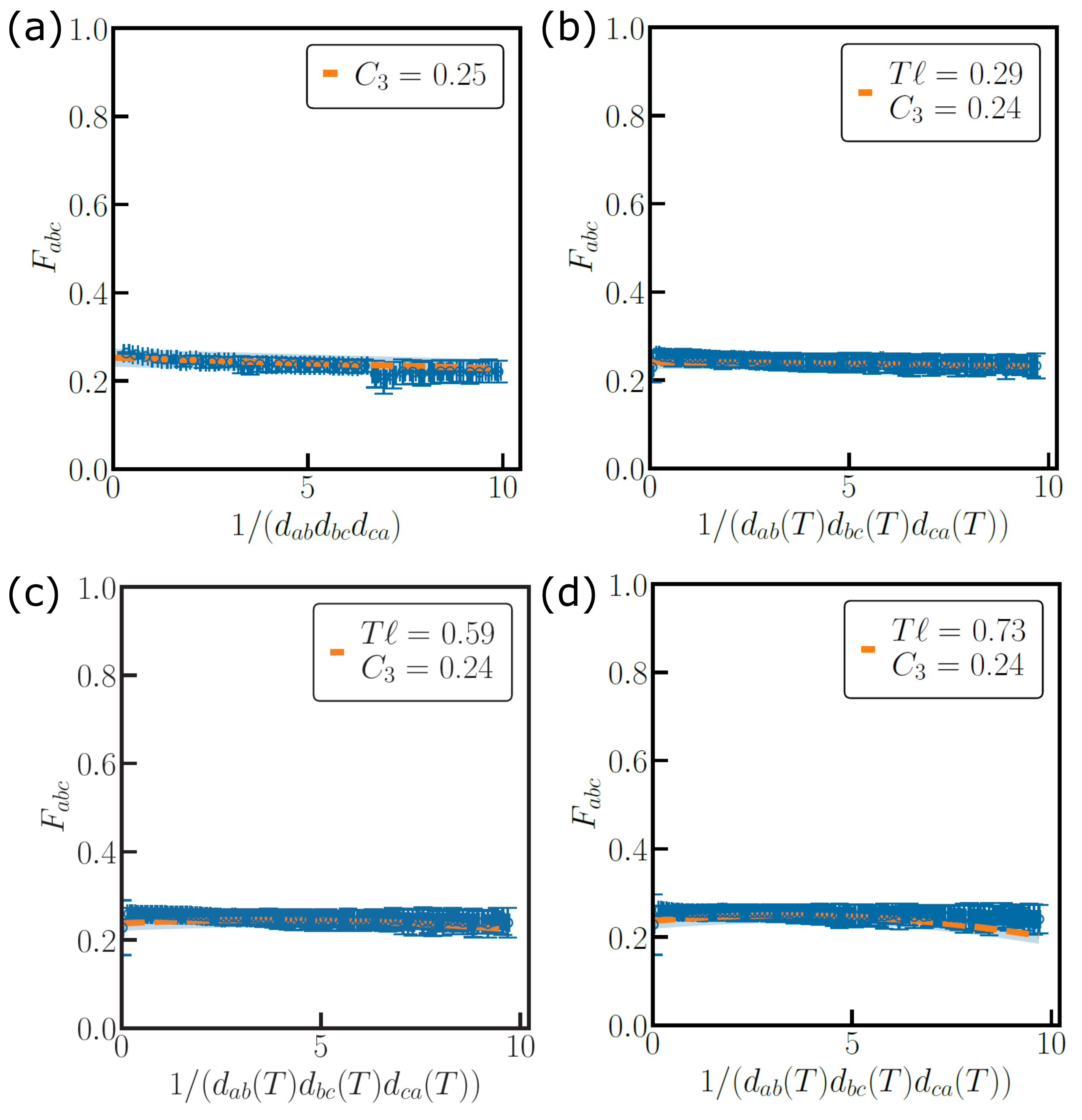}
  \caption{
    Asymptotic scaling form of the normalized ratio $F_{abc}$ for $m^2\ell^2=0.06$ 
    plotted in \textbf{(a)} for the 
    type-I boundary at zero-temperature, and for type-II boundary with $k=4$, $k=8$, and $k=10$ 
    \textbf{(b)-(d)}. The limit $F_{abc}\simeq C_3$ (left region of the plots where the argument approaches zero) is approximately independent of $T$, see Fig. 3e in the maintext.
  }
  \label{FabcComparison}
\end{figure}

As the comparison in Fig.~\ref{FabcComparison} shows for a fixed $m^2\ell^2$, 
this approach works satisfactory, with the long-distance
asymptotic behavior of $F_{abc}$ yielding an estimate of $C_3$. We find that $C_3$ remains consistently close across the type-I and type-II graphs. The coefficient
depends on $m^2\ell^2$ and $u$ as depicted in Fig.~3~(e) of the main text.

\section{S8. Simulation in electrical circuits}

We first describe how to realize the equation of motion (EOM) in Eq. (12) in an electrical circuit network consisting of resistors and diodes. Resistors $R_0$ are connected between circuit nodes according to the adjacency graph of the respective lattice, resistors $R_{\rm g}$ are connected between each node and ground, and a diode is placed between each node and ground. The resistors correspond to the linear terms of the EOM, whereas the diode introduces the on-site nonlinearity.

\begin{figure*}[t!]
  \includegraphics[width=0.9\linewidth]{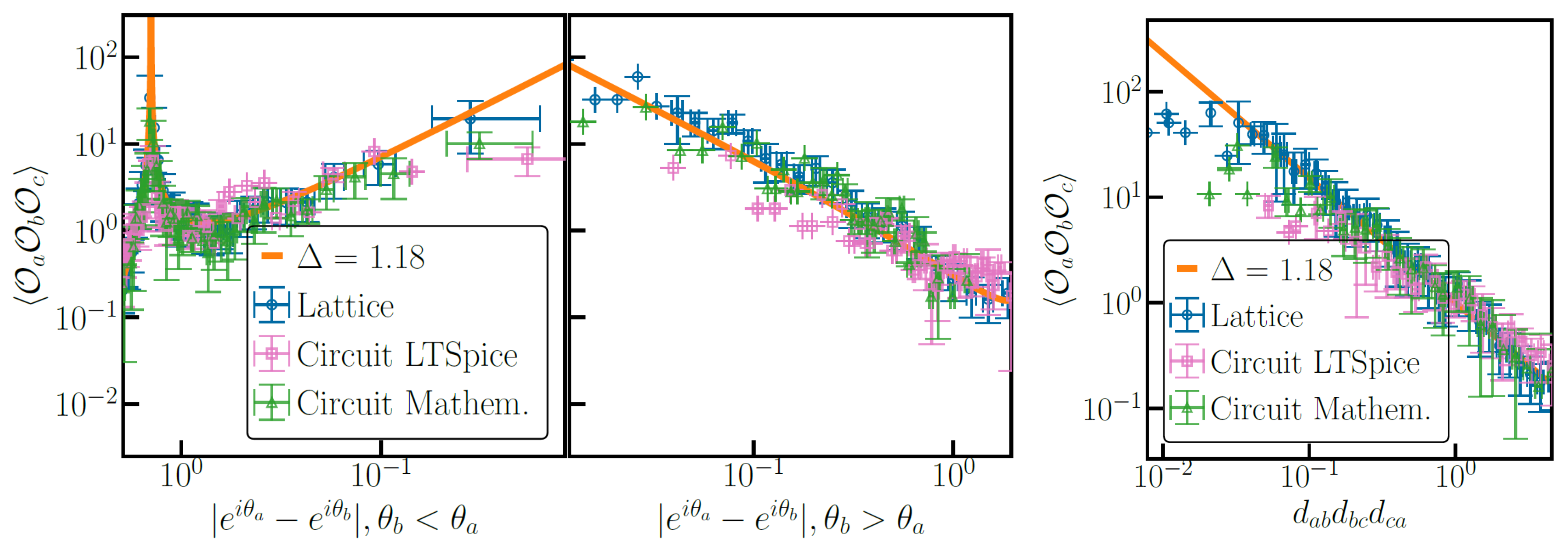}
  \caption{Comparison of circuit simulation using Mathematica (green) and LTspice (pink) for a small flake. For larger flakes with more sites, as presented in Fig. 3b of the main text, the LTspice simulation becomes noisier or larger distances, while the Mathematica simulation is still reliable. Note that the circuit data is normalized differently (choice of $C_3$) compared to Fig. 3b to make both data sets agree better for small distances.}
  \label{FigSim}
\end{figure*}

The current through a diode is approximately described by the Shockley equation $I(V) = I_\mathrm{S}(\mathrm{e}^{\frac{V}{V_{\rm S}}}-1)$, which, expanded to second order, reads $I(V) = c_1 V + c_2 V^2 + \mathcal{O}(V^3)$, with $c_1, c_2$ constants that depend on the specific diode. 
The values of $\bar{V}_\mu$ are represented by the voltages at the circuit nodes, with $\bar{V}_i$ in the bulk and $\bar{V}_a=J_a$ is realized by applying a fixed voltage at the respective boundary sites. The EOM only needs to be implemented for the bulk sites $\bar{V}_i$. Starting from the general EOM in Eq. (12) for any site $\mu$,
\begin{align}
\sum_\nu M_{\mu\nu}\bar{V}_\nu + \frac{u}{2}\bar{V}_\mu^2 = 0
\end{align}
with $M_{\mu\nu}=-A_{\mu\nu}+\hat{m}^2\delta_{\mu\nu}$, we find
\begin{align}
\sum_jM_{ij}\bar{V}_j + \frac{u}{2}\bar{V}_i^2 = I_i^{\rm eff}
\end{align}
for the bulk sites, with an effective bulk source term
\begin{align}
 I_i^{\rm eff} = I_i^{\rm eff}(\{J_a\}) = -\sum_a M_{ia}J_a.
\end{align}

To derive the EOM of the circuit, we add up all currents flowing out of one node, including up to second order in $V_i$. According to Kirchhoff's current law, we then obtain
\begin{align}
    I_i &= \frac{1}{R_0} \sum_j \left(q\, \delta_{ij}-A_{ij}\right) V_j +\frac{1}{R_{\rm g}} V_i + c_1 V_i + c_2 V_i^2.\label{EOMcirc}
\end{align}
Here, $I_i$ is any external current fed into the circuit nodes, for example from a connected voltage source. If no external supplies are added to a node then $I_i=0$, while a fixed voltage gives the corresponding current  $I_i=I_i^{\rm eff}(\{J_a\})$. \\
Furthermore, in practice (and in numerical simulations), a small parasitic capacitance will be present at each node, contributing an additional term of $C_{i} \dot{V}_i(t)$. Much like the process of discharging a capacitor, the voltages at the circuit nodes will decay exponentially towards the equilibrium state where $\dot{V}_i=0$, so that the solution for Eq. (12) is obtained as $t \to \infty$. In the following, we assume that all parasitic capacitances are so small that the equilibrium (or a sufficient approximation thereof) is reached instantaneously.

The relationship between circuit parameters and those of Eq. (12) is found from matching
\begin{align}
 &\sum_j(-A_{ij}+\hat{m}^2\delta_{ij})\bar{\phi}_j + \frac{u}{2}\bar{\phi}_i^2 \\
 &\stackrel{!}{=} c\Bigl[\frac{1}{R_0} \sum_j \left(q\, \delta_{ij}-A_{ij}\right) V_j +\frac{1}{R_{\rm g}} V_i + c_1 V_i + c_2 V_i^2\Bigr].
\end{align}
Here, $c=R_0/V_0$ is an arbitrary proportionality constant, and $V_0$ is a voltage scale that can be chosen freely (we set $V_0=1$ in the main text for simplicity). The matching yields the values shown this table:
\begin{center}
\begin{tabular}{|c|c|c|}
\hline
$\bar{\phi}_\mu$ & $\hat{m}^2$ & $u/2$\\
\hline
$V_\mu/V_0$ & $q+R_0(R_{\rm g}^{-1} + c_1)$ & $V_0R_0 c_2 $\\
\hline
\end{tabular}
\end{center}
We chose a Schottky diode (model RBE1VAM20A) for its low threshold voltage and high switching speed. From a calibration fit of the diode we extract the parameters $c_1=5.24 \cdot 10^{-4}$ and $c_2 = 9.43 \cdot 10^{-3}$. After arbitrarily fixing $R_0=100\Omega$, we choose $R_{\rm g}=240.3\Omega$ in order to obtain the target value of $m^2\ell^2 = \frac{1}{qh^2}R_0(R_{\rm g}^{-1}+c_1)= 0.271$, which is the mass used in Figs. 2 and 3 in the maintext. These parameters correspond to a non-linearity coefficient $u/2 = 0.943V_0$.

The linear-ramp protocol to compute $W_{\rm circ}(t)$ is implemented in the analog electronic circuit simulator computer software LTspice for each set$(a,b)$ and $(a,b,c)$ for the 2-pt and 3-pt correlation functions, respectively. The generating function of each run $W_{\rm circ}(t)$ is exported to a Mathematica notebook where the corresponding linear system is solved giving the correlation functions. For the 3-pt function, while the LTspice simulation works in practice, we use Mathematica to simulate the circuit EOM directly to reduce noise of the data. A comparison for small flakes, where the deviations are smaller, is shown in Fig. \ref{FigSim}.

\vfill

\end{document}